\documentclass{natureprintstyle}
\usepackage{amssymb,amsmath,graphicx,fullpage}

\usepackage[square,comma,numbers]{natbib}
\usepackage{pdfpages}

\usepackage[left=1.8cm,top=2cm,right=1.8cm,bottom=1.8cm,nohead]{geometry}

\newcommand\arcmin{\mbox{$^\prime$}}%
\newcommand\arcsec{\mbox{$^{\prime\prime}$}}%

\newcommand\snfe{\mbox{SN\,2011fe}}
\newcommand\snptf{\mbox{PTF11kly}}

\def\simlt{\mathrel{\hbox{\rlap{\hbox{\lower4pt\hbox{$\sim$}}}\hbox{$<$}}}}
\def\simgt{\mathrel{\hbox{\rlap{\hbox{\lower4pt\hbox{$\sim$}}}\hbox{$>$}}}}

\def\ale{\mathrel{\hbox{\rlap{\hbox{\lower4pt\hbox{$\sim$}}}\hbox{$<$}}}}
\def\age{\mathrel{\hbox{\rlap{\hbox{\lower4pt\hbox{$\sim$}}}\hbox{$>$}}}}

\def\aj{{\rm AJ}}                   
\def\araa{{\rm ARA\&A}}             
\def\apj{{\rm ApJ}}                 
\def\apjl{{\rm ApJ}}                
\def\apjs{{\rm ApJS}}

\def\aap{{\rm A\&A}}

\def\mnras{{\rm MNRAS}}             
\def\nar{{\rm New~Ast.~Rev.}}

\def\pasp{{\rm PASP}}

\def\spose#1{\hbox to 0pt{#1\hss}}
\newcommand\lsim{\mathrel{\spose{\lower 3pt\hbox{$\mathchar"218$}}
     \raise 2.0pt\hbox{$\mathchar"13C$}}}
\newcommand\gsim{\mathrel{\spose{\lower 3pt\hbox{$\mathchar"218$}}
     \raise 2.0pt\hbox{$\mathchar"13E$}}}

\usepackage{times}
\usepackage{mathptmx}

\begin{document}
	\title{Constraints on the Progenitor System\\
	of the Type Ia Supernova\\
	SN\,2011fe/PTF11kly}
	
\author{Weidong~Li$^{1}$,
		Joshua~S.~Bloom$^{1}$,
		Philipp~Podsiadlowski$^{2}$,
		Adam~A.~Miller$^{1}$,
		S.~Bradley~Cenko$^{1}$,
		Saurabh~W.~Jha$^{3}$,
		Mark~Sullivan$^{2}$,
		D.~Andrew~Howell$^{4,5}$,
		Peter~E.~Nugent$^{6,1}$,
		Nathaniel R. Butler$^{7}$,
		Eran~O.~Ofek$^{8,9}$,
		Mansi~M.~Kasliwal$^{10}$,
		Joseph~W.~Richards$^{1,11}$,
		Alan~Stockton$^{12}$,
		Hsin-Yi~Shih$^{12}$,
		Lars~Bildsten$^{5,13}$,
		Michael~M.~Shara$^{14}$,
		Joanne Bibby$^{14}$,
		Alexei~V.~Filippenko$^{1}$,
		Mohan~Ganeshalingam$^{1}$,
		Jeffrey~M.~Silverman$^{1}$,
		S.~R.~Kulkarni$^{8}$,
		Nicholas~M.~Law$^{15}$,
		Dovi~Poznanski$^{16}$,
		Robert~M.~Quimby$^{8}$,
		Curtis~McCully$^{3}$,
		Brandon~Patel$^{3}$,
		\& Kate~Maguire$^{2}$
}
\maketitle

\begin{affiliations}
	\item Department of Astronomy, University of California, Berkeley, CA	
	      94720-3411, USA. 
	\item Department of Physics (Astrophysics), University of Oxford,
	      Keble Road, Oxford OX1 3RH, UK. 
	\item Department of Physics and Astronomy, Rutgers, The State 
	      University of New Jersey, Piscataway, NJ 08854, USA. 
	\item Las Cumbres Observatory Global Telescope Network, Goleta, CA 
	      93117, USA. 
	\item Department of Physics, University of California, Santa Barbara,
	      CA 93106, USA. 
	\item Computational Cosmology Center, Lawrence Berkeley National
	      Laboratory, 1 Cyclotron Road, Berkeley, CA 94720, USA. 
	\item Department of Physics, Arizona State University, Tempe, AZ 85287-1504, USA. 
	\item Cahill Center for Astrophysics 249-17, California Institute of
	      Technology, Pasadena, CA, 91125, USA. 
	\item Benoziyo Center for Astrophysics, Faculty of Physics, Weizmann
		Institute of Science, 76100 Rehovot, Israel.
	\item Carnegie Institution for Science, 813 Santa Barbara Street, Pasadena CA 91101, USA. 
	\item Department of Statistics, University of California, Berkeley, CA
		      94720-7450, USA. 
	\item Institute for Astronomy, University of Hawaii, Honolulu, HI 96822,
	      USA. 
	\item Kavli Institute for Theoretical Physics, University of California, Santa Barbara, CA 93106, USA. 
	\item Department of Astrophysics, American Museum of Natural 
		      History, Central Park West and 79th street, New York, NY 10024-5192,
		      USA. 
	\item Dunlap Institute for Astronomy and Astrophysics, University of
	      Toronto, 50 St. George Street, Toronto M5S 3H4, Ontario, Canada. 
	\item School of Physics and Astronomy, Tel-Aviv University, Tel-Aviv 69978, Israel.
\end{affiliations}

	\date{\today}{}

\begin{abstract}
\phantom{}\vskip 0.05cm 
Type Ia supernovae (SNe) serve as a fundamental pillar of modern cosmology, owing to their large luminosity and a well-defined relationship between light-curve shape and peak brightness\cite{1993ApJ...413L.105P,2011NatCo...2E.350H}. The precision distance measurements enabled by SNe~Ia first revealed the accelerating expansion of the universe\cite{1998AJ....116.1009R,1999ApJ...517..565P}, now widely believed (though hardly understood) to require the presence of a mysterious ``dark'' energy. General consensus holds that Type Ia SNe result from thermonuclear explosions of a white dwarf (WD) in a binary system\cite{1982ApJ...253..798N,1984ApJS...54..335I}; however, little is known of the precise nature of the companion star and the physical properties of the progenitor system.
Here we make use of extensive historical imaging obtained
at the location of \snfe/\snptf, the closest SN~Ia discovered
in the digital imaging era, to constrain the visible-light luminosity of the progenitor to be 10--100 times fainter than previous limits on other SN~Ia progenitors. This directly
rules out luminous red giants and the vast majority of helium stars as the mass-donating companion to the exploding white dwarf.  Any evolved red companion must have been born with mass less than 3.5 times the mass of the Sun.
These observations favour a scenario where the exploding WD of \snfe/\snptf\, 
accreted matter either from another WD, or by Roche-lobe overflow from a
subgiant or main-sequence companion star. 

\end{abstract}


\phantom{}\vskip 0.05cm \noindent 

\snfe/\snptf\, (hereafter SN\,2011fe) was discovered in the Pinwheel Galaxy (Messier 101; M101), a spiral galaxy at a distance\cite{1998ApJ...508..491S,2011ApJ...733..124S} of 6.4 Megaparsec (Mpc), by the Palomar Transient Factory (PTF) collaboration on 2011 Aug.\ 24 UT\cite{thenuge}. The galaxy was intensively monitored by our collaboration over the past decade and the site of \snfe\ was fortuitously observed prior to the explosion by the {\it Hubble Space Telescope (HST)} on several occasions.
Together, these archival data offer a unique opportunity to constrain the nature
of the progenitor system of \snfe.

The absence of hydrogen and helium, coupled with the presence of silicon
in the spectra of SNe~Ia, are the distinguishing signatures of the class\cite{1997ARA&A..35..309F}
and exclude most massive stars as SN~Ia progenitors. This, taken
with the significant inferred amount of synthesised  radioactive $^{56}$Ni
(0.1--0.9 M$_\odot$)\cite{2007Sci...315..825M} created in
the explosions, suggests that the majority of SNe~Ia
arise from rapid thermonuclear fusion of a carbon-oxygen
white dwarf (WD) that is more massive than that made from single-star evolution (greater than 1
solar mass).
A more massive WD also has the higher densities needed to
ignite the runaway carbon fusion reaction and trigger the observed nucleosynthesis.
Thus, a companion star that donates mass to the WD is required, even though
fundamental questions remain as to how the accretion of matter leads to
the explosion for each progenitor model.

The broadly diverse progenitor models are organised into two classes: double degenerate (DD) and single degenerate (SD)\cite{1997Sci...276.1378N,2008NewAR..52..381P}.
A DD model involves two WDs in a close binary system\cite{1984ApJS...54..335I,1984ApJ...277..355W}; due to the release of gravitational radiation, the orbit shrinks. Ultimately, the lighter object (secondary) is disrupted and accretion onto the primary ignites runaway thermonuclear fusion. Alternatively, in SD models\cite{1982ApJ...253..798N,1973ApJ...186.1007W}, the primary WD accretes material from a stellar companion until its mass approaches the Chandrasekhar mass, $M_{\rm Ch} \approx 1.4$ M$_\odot$, at which point again a thermonuclear explosion ensues. Different SD models can be distinguished based on the nature of the secondary: the WD can accrete either from 
a wind (the ``symbiotic channel"\cite{1992ApJ...397L..87M}), by Roche-lobe overflow (the
``RLOF channel"\cite{1992A&A...262...97V}), or by mass transfer from a helium star
(the ``helium star channel"\cite{1982ApJ...253..798N,2010A&A...523A...3L}). The secondary star
in the symbiotic channel is often a red-giant star, while it is a subgiant or main-sequence
star in the RLOF channel. 

Because WDs are faint and can only be observed directly in our own Milky Way 
and several very nearby galaxies, much of the focus in SN~Ia 
progenitor searches has been on determining the nature of the
companion star. Previous efforts (see Supplementary Table 1) have achieved an absolute magnitude in the $V$ band of $M_V \approx -5$ mag, a limit that is not deep enough to rule out any plausible progenitor systems. 

We have analysed the available {\it HST} observations of M101 (see Supplementary Information for details). To pinpoint the precise SN location in the {\it HST} images, we obtained 
a mosaic image of the field of \snfe\ with the Near-Infrared Camera 2
(NIRC2) mounted behind the adaptive optics (AO) system on the Keck II
telescope\cite{wlb+06}. The Keck AO image was astrometrically registered to the {\it HST}/ACS images, yielding a $1\sigma$ precision of 0$\arcsec$.021 (or 21 mas) for the SN position (see Supplementary Information). Figure 1 shows the site of \snfe\ on different scales. 
No object is detected at the nominal SN location in 4 different {\it HST} filters, or within the $\sim 8\sigma$ error radius.

This non-detection can be directly translated into limits on the progenitor system brightness and temperature. Indeed, each of the four {\it HST} bands place different limits on the progenitor brightness depending on the (unknown) effective temperature and luminosity class. These limits in turn can be converted to an equivalent $V$-band magnitude, allowing us to quote the most restrictive single-frame image limit or a limit derived from a temperature-dependent stack of all images; these quantities, as well as the absolute-magnitude limits at the distance of M101, are given in the Supplementary Information.

Figure 2 shows the region in effective temperature vs. absolute $V$-band magnitude excluded by
the {\it HST} imaging analysis. At 3000 K, progenitor systems of SN\,2011fe are excluded for $M_V
\le 1$ mag, and for temperatures larger than $T_{\rm eff} \approx 5000$ K, the exclusion is $M_V
\le -0.5$ mag. These limits rule out a
symbiotic binary progenitor (a red giant and high-mass WD) such as RS~Oph and
probably T~CrB\cite{2001ApJ...558..323H}. Similarly, the helium-star binary V445~Pup\cite{2008ApJ...684.1366K} is
only marginally consistent with the upper limits, although the entire He-star channel 
cannot be completely excluded (see the blue-shaded 
area in Figure 2).  Among the SD models,
only the RLOF channel (such as U~Sco\cite{2001MNRAS.327.1323T}) can easily be
reconciled with the {\it HST} constraints. Finally, all variations of the DD model are consistent with the 
non-detection of a source at the position of \snfe, as it does not
predict a bright source in the visible range. The brightness limits we have
deduced also rule out a globular cluster, or any open star cluster with more
than $\sim 300$ members as the site of \snfe. 

Figure 2 also shows that a mass donor with an effective temperature less
than 4800 K (spectral type redder than $\sim$ G5) needs to have a zero-age main
sequence (ZAMS) mass less than 2.2 M$_\odot$. For stars with spectral type redder
than an A0 star (effective temperature $\sim$ 10,000 K), the companion star would
need to be less massive than 3.5 M$_\odot$.
For $T_{\rm eff} \approx 3000$--4000~K, as expected for the red-giant branch (RGB) stars, the $M_V$
limit excludes progenitors brighter than an absolute $I$-band mag $M_I \approx -2$. This
limit is 2 mag fainter than the observed\cite{2011ApJ...733..124S} tip of the RGB in M101 and
places an upper bound to the radius of $R \ale 60~ {\rm R}_\odot$ for $T_{\rm eff} = 3500 {\rm K}$ on any RGB
progenitor. In a progenitor model that requires RLOF, this limit then demands an orbital period smaller
than $P_{\rm orb}=260$ to 130 days in a binary system with a 1.3 M$_\odot$ WD (where the range of
$P_{\rm orb}$ accommodates the 0.5--2.5 M$_\odot$ range allowed for an RGB star).

RS~Oph, T~CrB, and U~Sco are recurrent novae in the Milky Way. The recurrence time of WD
binaries where the WDs are close to the Chandrasekhar mass is
expected\cite{2010ApJS..187..275S} to be 10--20\,yr. Though our historical images had sufficient sensitivity to detect classical novae, we find no evidence for any such outburst in the past 12\,yr  at the site of \snfe. However, we estimate a $\sim$37\% chance that a typical nova could have occurred in the past $\sim$5 yr and have been missed given the particular cadence of the imaging (see Supplementary Information).

Historical imaging at other wavebands complements these visible-light 
progenitor system constraints.  We have analysed 11 epochs
of archival {\it Chandra} X-ray observations of M101 taken in 2004 (see
Supplementary Information), and derived upper limits for the X-ray
luminosity at the location of SN\,2011fe in the range (4--25) $\times 10^{36}$\,erg\,s$^{-1}$
(depending on the details of the assumed spectrum). 
SD progenitor systems are thought to undergo a prolonged period ($\Delta t \approx
10^{6}$\,yr) of steady nuclear burning during the mass-transfer process.  Such systems
should appear as luminous [$L_{\rm X} \approx 10^{36}$--$10^{38}$\,erg\,s$^{-1}$ X-ray ($kT \approx 100$\,eV)]
sources, and indeed nearly 100 of these super-soft sources (SSSs) have been identified to date
in the Milky Way and other nearby galaxies, including M101 itself\cite{g00,d10a}.
DD progenitor systems have also been predicted to emit X-rays \cite{d10b,ypr07}, with 
$L_{X} \approx 10^{36}$--$10^{37}$\,erg\,s$^{-1}$. Our historical
X-ray limits are, unfortunately, not deep enough to rule out either channel. 

We have also analysed pre-explosion {\it Spitzer} mid-infrared (IR) data of M101 taken
in 2004 (see Supplementary Information). No point source was detected
at the SN location, to a mid-IR luminosity of $<$ (1--13) $\times 10^{36}~ {\rm erg\,s}^{-1}$.
These limits rule out bright red-giant mass donors 
of  \snfe, in the symbiotic channel, but are consistent with a relatively faint 
secondary star in the RLOF channel. 
For a DD merger to produce a SN~Ia, the merger product may need to
evolve and cool slowly\cite{2011arXiv1108.4036S} for $> 1\times10^6$ yr. 
This requires the radiated luminosity (mostly in the IR) to be
$< 4 \times 10^{36}~ {\rm erg\,s}^{-1}$. 
The {\it Spitzer} non-detection limits are certainly consistent
with the requirement of low IR luminosity for the DD models. 

One test to determine whether SN\,2011fe comes from the RLOF
channel of the SD model, or from a DD merger, is to detect the 
surviving companion in the RLOF channel. But with our current
resources, this is a very challenging observation for a
subgiant mass donor, and nearly impossible for a main-sequence companion,
unless there is excessive heating caused by the interaction between
the companion star and the SN ejecta.

The observations of SN\,2011fe and historical imaging at the 
SN site have directly excluded symbiotic binary
and probably helium-star progenitors. This, along with the 
X-ray and mid-IR limits,  suggest that the WD exploded as 
SN\,2011fe was accreting matter either from another WD, or by RLOF from a subgiant or main-sequence star. 
Given the observed diversity of SNe~Ia, however, the possibility of
more than one progenitor  for the class remains. As the trove of high-resolution wide-field
imaging grows, covering more and more future SN explosion 
sites, similar analyses will eventually shed light on this possibility.

\begin{addendum}

 \item[Correspondence] Correspondence and requests for materials
should be addressed to WL\ \\(email: weidong@berkeley.edu).

 \item[Competing Interests] The authors declare that they have no
competing financial interests.

\item[Acknowledgements] The authors thank K. Shen for helpful
  discussions and to the staff of W. M. Keck Observatory, especially
  J.\ Lyke and R.\ Campbell, for their assistance in helping obtain
  the NIRC AO imaging.  JSB, AAM, and JWR were partially supported by
  an NSF-CDI grant (award \#0941742) ``Real-time Classification of
  Massive Time-series Data Streams.'' PP acknowledges very helpful
  discussions on symbiotic binaries with J. Miko\l ajewska. MMK
  acknowledges support by NASA's Hubble Fellowship and the
  Carnegie-Princeton Fellowship. AVF's group at UC Berkeley has
  received financial assistance from Gary \& Cynthia Bengier, the
  Richard \& Rhoda Goldman Fund, NASA/{\it HST} grant AR-12126 from
  the Space Telescope Science Institute (which is operated by AURA,
  Inc., under NASA contract NAS 5-26555), the TABASGO Foundation, and
  NSF grant AST-0908886.  EOO is supported by an Einstein Fellowship
  from NASA.  MMS acknowledges the support of Hilary Lipsitz and the
  American Museum of Natural History for essential funding. MS
  acknowledges support from the Royal Society. Support for this
  research at Rutgers University was provided in part by NSF CAREER
  award AST-0847157 to SWJ. LB acknowledges support from NSF grant
  AST-1109174.  Some of the data presented herein were obtained at the
  W. M. Keck Observatory, which is operated as a scientific
  partnership among the California Institute of Technology, the
  University of California, and NASA; the observatory was made
  possible by the generous financial support of the W. M. Keck
  Foundation.  Observations were obtained with the Samuel Oschin
  Telescope at the Palomar Observatory as part of the Palomar
  Transient Factory project, a scientific collaboration between the
  California Institute of Technology, Columbia University, La Cumbres
  Observatory, the Lawrence Berkeley National Laboratory, the National
  Energy Research Scientific Computing Center, the University of
  Oxford, and the Weizmann Institute of Science. The National Energy
  Research Scientific Computing Center, which is supported by the
  Office of Science of the U.S. Department of Energy under Contract
  No.\, DE-AC02-05CH11231, provided staff, computational resources,
  and data storage for this project.

 \item[Author Contributions] WL, JSB, SWJ, CM, and BP analysed the
   {\it HST} photometry in the context of progenitor limits. PP
   contributed analysis of progenitor models. AAM, JWR, and SBC
   analysed historical imaging from PTF and KAIT in the context of
   novae limits. MMK provided analysis of {\it Spitzer}
   observations. MMS and JB provided analysis of the {\it HST}
   imaging. MMS also contributed interpretation of the progenitor
   limits. NRB, EOO, and LB contributed analysis and interpretation of
   the historical X-ray imaging. DP, RMQ, SRK, NML, EOO, SBC, MS, DAH,
   JSB, PEN, MMK, LB, and KM were responsible for obtaining, reducing,
   and analysing the PTF observations. AS and HYS obtained the Keck AO
   imaging and SBC reduced and analysed those images. AVF, MG, WL, and
   JMS were responsible for the KAIT imaging and analysis. AVF also edited
   the final manuscript.
\end{addendum}

\clearpage 

\begin{figure}
	\label{fig:color}
       \centering
        \parbox{6.5in}{
      \includegraphics[width=2.2in,angle=270]{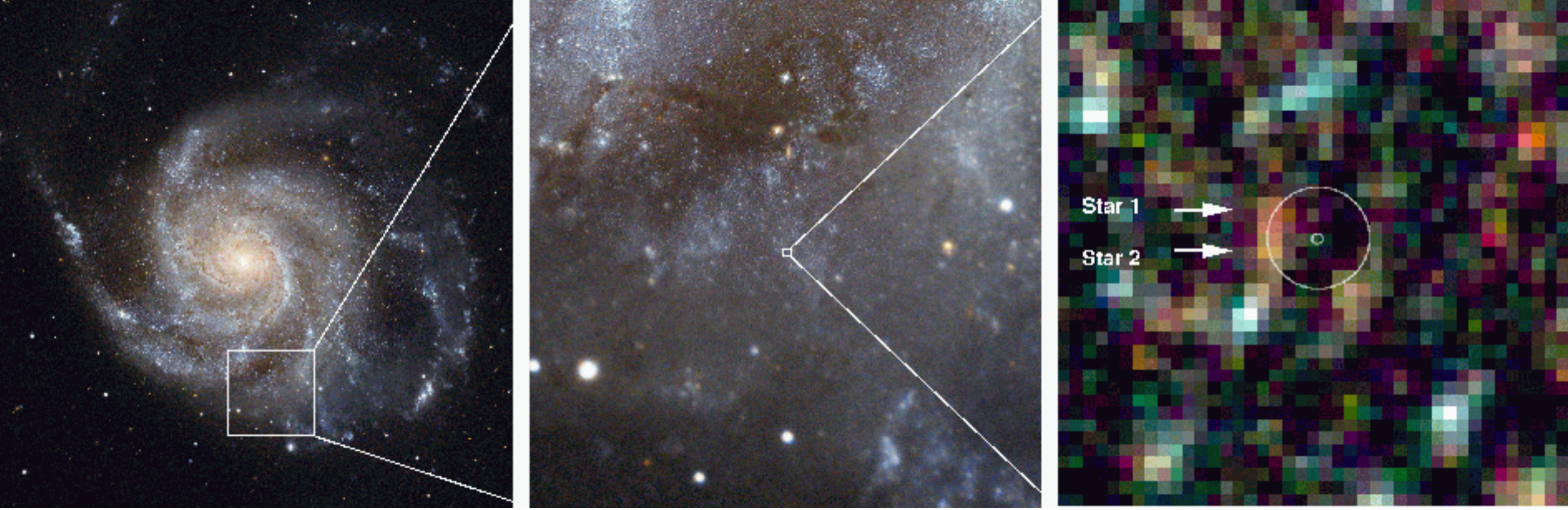}
        \centering\parbox{6.5in}{
\caption{\textbf{The site of \snfe\ in Messier 101 as imaged by HST/ACS.}  		The left panel is a full-view colour picture of the face-on spiral galaxy M101 ($18\arcmin \times 18\arcmin$ field of view) constructed from the three-colour {\it HST}/ACS images taken at multiple mosaic pointings (from {\tt http://hubblesite.org}). North is up and east to the left. M101 displays
		several well-defined spiral arms. With a diameter of 170 thousand light years across,
		M101 is nearly twice the size of our Milky Way Galaxy, and is estimated to contain
		at least one trillion stars. The middle panel is a cutout section 
		($3\arcmin \times 3\arcmin$) of the left panel, centred on the SN location.
		\snfe\ is spatially projected on a prominent spiral arm. The right panel
		is a section of $2\arcsec \times 2\arcsec$ centred on the SN location,
		which is marked by two circles. The smaller circle has a radius of our
		$1\sigma$ astrometric uncertainty (21 mas), while the bigger circle has a
		radius of 9 times that. No object is detected at the nominal SN location,
		or within the $8 \sigma $ error radius. Two nearby, but unrelated, red sources
		are labeled as ``Star 1" and ``Star 2," and are displaced from our nominal SN location by $\sim 9\sigma$, formally excluded as viable candidate objects involved in the progenitor system of \snfe. Credit for the left panel colour picture:
		NASA, ESA, K. Kuntz (JHU), F. Bresolin (University of Hawaii),
		J. Trauger (Jet Propulsion Lab), J. Mould (NOAO), Y.-H. Chu
		(University of Illinois, Urbana), and STScI. {\it Note: This is a reduced-size figure for arxiv posting.}}}}
\end{figure}
\clearpage 

\begin{figure}
	\label{fig:closeup}
       \centering
        \parbox{3.6in}{
       \includegraphics[width=4.5in,angle=0]{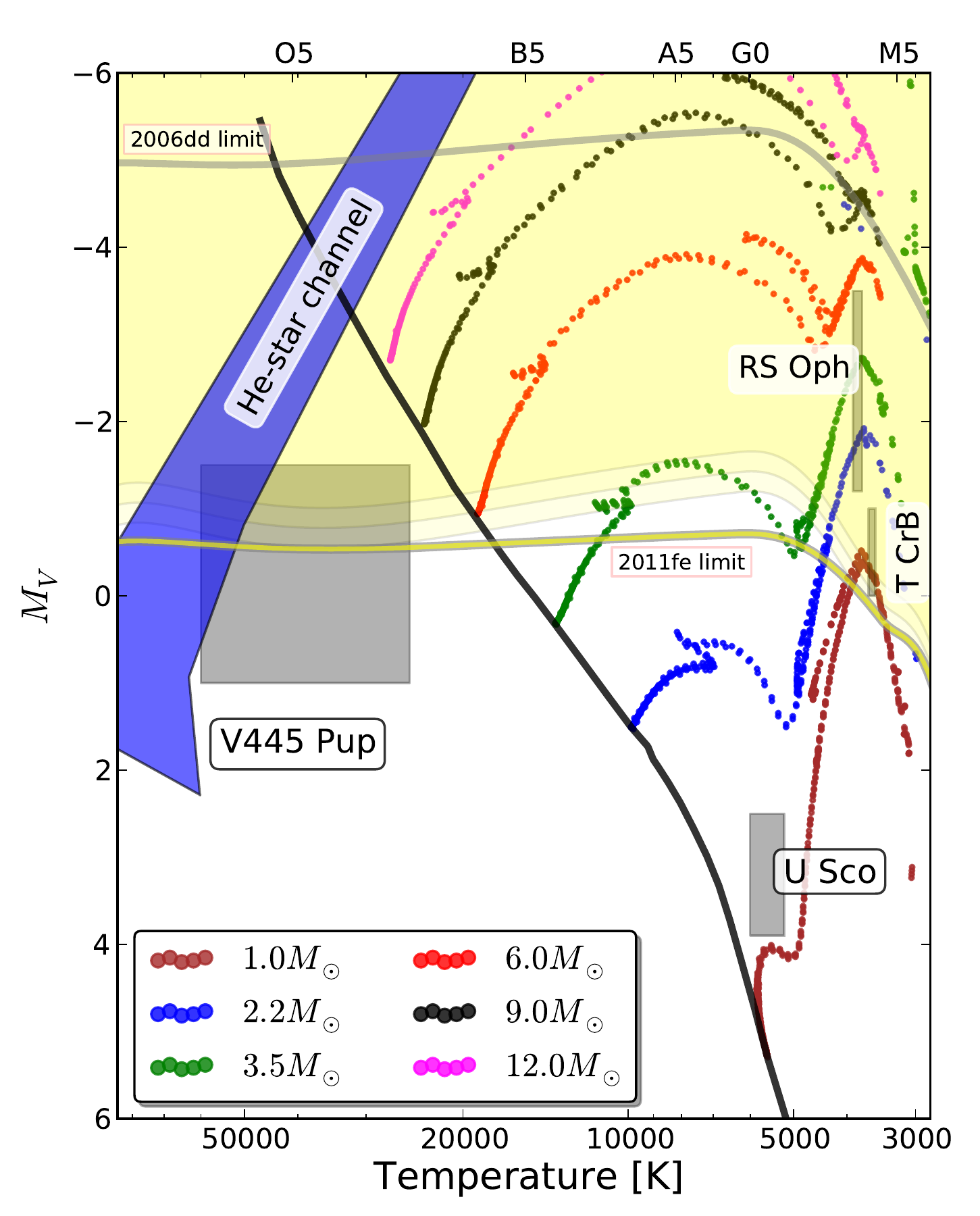}
        \centering\parbox{3.6in}{
\caption{\textbf{Progenitor system constraints in a Hertzsprung-Russell (H-R) diagram compared to some proposed single-degenerate progenitors.} The thick yellow line is the 2$\sigma$ limit in $V$-band absolute magnitude ($M_V$) against effective temperature at the SN location (see text) from a combination of the four {\it HST} filters, weighted using synthetic colours of redshifted stellar spectra at solar metallicity for that temperature and luminosity class. A more conservative limit comes from taking the single filter that most
constrains the stellar type and luminosity class; shown is the 2$\sigma$ limit assuming
the adopted distance modulus\cite{1998ApJ...508..491S,2011ApJ...733..124S} of 29.04\,mag (middle light yellow curve) with a total uncertainty of 0.23 mag (top/bottom light yellow curve). Depicted are the theoretical estimates (He-star channel\cite{2010A&A...523A...3L}) and observed candidate systems (V445~Pup\cite{2008ApJ...684.1366K}, RS~Oph\cite{2001ApJ...558..323H}, U~Sco\cite{1999ApJ...519..314H,2001MNRAS.327.1323T}, and T~CrB\cite{2001ApJ...558..323H}). Also plotted are theoretical
		evolutionary tracks (from 1 Myr to 13 Gyr) of isolated stars for a range of masses
		for solar metallicity; note that the limits on the progenitor mass of SN\,2011fe
		under the supersolar metallicity assumption are similar to those represented here. The grey
		curve at top is the limit inferred from {\it HST} analysis of SN\,2006dd, representative of the other nearby SN~Ia progenitor limits (see Supplementary Information). For the helium-star channel, bolometric luminosity corrections to the $V$ band are adopted based on effective temperature\cite{2010AJ....140.1158T}. The foreground Galactic and M101 extinction due to dust is negligible\cite{thenuge} and taken to be $A_V = 0$ mag here. Had a source at the $2.0\,\sigma$ photometric level been detected in the {\it HST} images at the precise location of the SN, we would have been able to rule out the null hypothesis of no significant progenitor with 95\% confidence. As such, we use the 2$\sigma$ photometric uncertainties in quoting the brightness limits on the progenitor system.}}}

\end{figure}



\section*{Supplementary material}

# Constraints on the Progenitor of the Type Ia Supernova PTF11kly/SN 2011fe.

## Supplementary Methods

### Prior limits on SN Ia progenitors from optical data

Nine SNe Ia with preexisting *HST* data of their host galaxy have been close enough (within 25 Mpc) to search for a progenitor (Supplementary Table 1). No progenitor has been found; only upper limits have been possible[1,2,3,4]. Limits range from $M_g = -3.9$ mag in the case of SN 2004W to $M_I = -8.3$ mag for SN 2003cg[1]. These limits provided only poor constraints — in the case of SN 2006dd and SN 2006mr in NGC 1316, ruling out normal stars with initial masses greater than 6 $M_\odot$ at the tip of the asymptotic giant branch (AGB), young post-AGB stars with initial masses greater than 4 $M_\odot$, and post-red-giant stars with initial masses greater than 9 $M_\odot$[2].

## HST data reduction

### Search for the progenitor system of SN 2011fe

We searched the *HST* data archive and found six datasets that contain the site of SN 2011fe. These include the ACS images observed in the F435W, F555W, and F814W filters in program GO–9490 (PI: K. Kuntz) on 2002 Nov. 13, a pair of images observed with the Wide Field Planetary Camera 2 (WFPC2) in the F606W filter in program GO–7909 (PI: S. Casertano) on 1998 May 22, and a Wide Field Camera 3 (WFC3) image observed in the F469N filter in program GO–11635 (PI: M. Shara) on 2010 Apr. 7. More details on the available *HST* data are listed in Supplementary Table 2.

    To initially locate the site of PTF11kly in these images, we utilised the PTF discovery images reported by Nugent et al. (2011a)[5]. We followed the general procedure to identify SN progenitors in high-resolution *HST* images as reported by, for example, Li et al.(2005)[6]. We first registered the PTF images to a deep *r*-band MegaCam image downloaded from the 3.6-m Canada-France-Hawaii Telescope (CFHT) science data archive[1]. 12 stars were used in the astrometric solution for this step, yielding a precision of $1\sigma = 0''.055$. The CFHT MegaCam image was further registered to the ACS/F814W image. The astrometric solution, using 16 stars, has a precision of $1\sigma = 0''.017$. The $1\sigma$ uncertainty of the final SN location in the ACS/F814W image is $0''.058 = 58$ mas. No object is visible at the nominal position of SN 2011fe in any of the *HST* images. This finding was reported as an astronomer's telegram (ATEL; Li et al. 2011a)[7]. We noted in our ATEL the presence of two red sources within roughly the $2 \times 1\sigma$ uncertainty of our solution. Further analysis of these two sources was reported by Li et al. (2011b)[8].

    To improve our astrometric solution, on 2011 August 26 we observed the field of PTF11kly with the Near-Infrared Camera 2 (NIRC2) mounted behind the adaptive optics (AO) system on the Keck II telescope[9]. Beginning at 5:33, we obtained a series of *H*-band exposures, each consisting of 30 nondestructive readouts of 2 s each, for a total time on source of 360 s. The SN was sufficiently bright to use as the guide star for the tip-tilt correction. Images were reduced using standard IRAF[2] routines, using a median combination of the (nonaligned) dithered images to correct

---

[1] This research used the facilities of the Canadian Astronomy Data Centre operated by the National Research Council of Canada with the support of the Canadian Space Agency.

[2] IRAF is distributed by the National Optical Astronomy Observatory, which is operated by the Association of Universities for Research in Astronomy (AURA), Inc., under cooperative agreement with the National Science Foundation (NSF).

| Supernova | Distance (Mpc) | Limits (mag)[a] | | | Reference |
|---|---|---|---|---|---|
| SN 2003cg | 17.3 | $M_R = -7.1$ | $M_I = -8.3$ | | 1 |
| SN 2003gs | 17.9 | $M_V = -6.2$ | | | 1 |
| SN 2003hv | 21.3 | $M_V = -6.3$ | $M_I = -6.8$ | | 1 |
| SN 2004W | 17.3 | $M_g = -3.9$ | $M_Z = -5.5$ | | 1 |
| SN 2005df | 24 | $M_B = -6.2$ | $M_R = -5.1$ | | 1 |
| SN 2006dd[c] | 19 | $M_{B*} = -5.4$ | $M_V = -5.3$ | $M_I = -6.0$ | 2 |
| SN 2006mr[c] | 19 | $M_{B*} = -5.6$ | $M_V = -5.4$ | $M_I = -6.0$ | 2 |
| SN 2007sr | 12–25 | $M_V = -5.5$ to $-4.1$ | $M_I = -7.5$ to $-5.6$ | | 3 |
| SN 2007on | 20 | $M_g = -4.5$ | $M_I = -4.5$ | | 4 |

**Supplementary Table 1** — Limits on SN Ia progenitors from preexisting *HST* data. [a]Filter Key: $B$ = F450W, $B_*$ = F435W, $g$ = F475W, $V$ = 555W, $R$ = F606W, $I$ = F814W, $Z$ = F850LP. Photometric system is AB for ref. 2, Vega for ref. 4. Others are not stated. [b]SN 2004ab at 14.1 Mpc has no constraints due to the high surface brightness in the galactic bulge. [c]Both SN 2006dd and SN 2006mr were located in the same galaxy, NGC 1316.

| Dataset | Date(obs) | Exptime(s) | Instrument | Aperture | Filter | Prop. No. | Label[a] |
|---|---|---|---|---|---|---|---|
| J8D6A2011 | 2002-11-13 | 900.0 | ACS | WFC | F435W | 9490 | ACS/F435W |
| J8D6A2021 | 2002-11-13 | 720.0 | ACS | WFC | F555W | 9490 | ACS/F555W |
| J8D6A2031 | 2002-11-13 | 720.0 | ACS | WFC | F814W | 9490 | ACS/F814W |
| U4K2KF02R | 1998-05-22 | 260.0 | WFPC2 | WFALL | F606W | 7909 | WFPC2/F606W |
| U4K2KF01R | 1998-05-22 | 260.0 | WFPC2 | WFALL | F606W | 7909 | WFPC2/F606W |
| IB3P17010 | 2010-04-08 | 6106.0 | WFC3 | UVIS | F469N | 11635 | WFC3/F469N |

**Supplementary Table 2** — Available *HST* data at the location of PTF11kly/SN 2011fe.
[a]These labels are used in the text to represent the images.

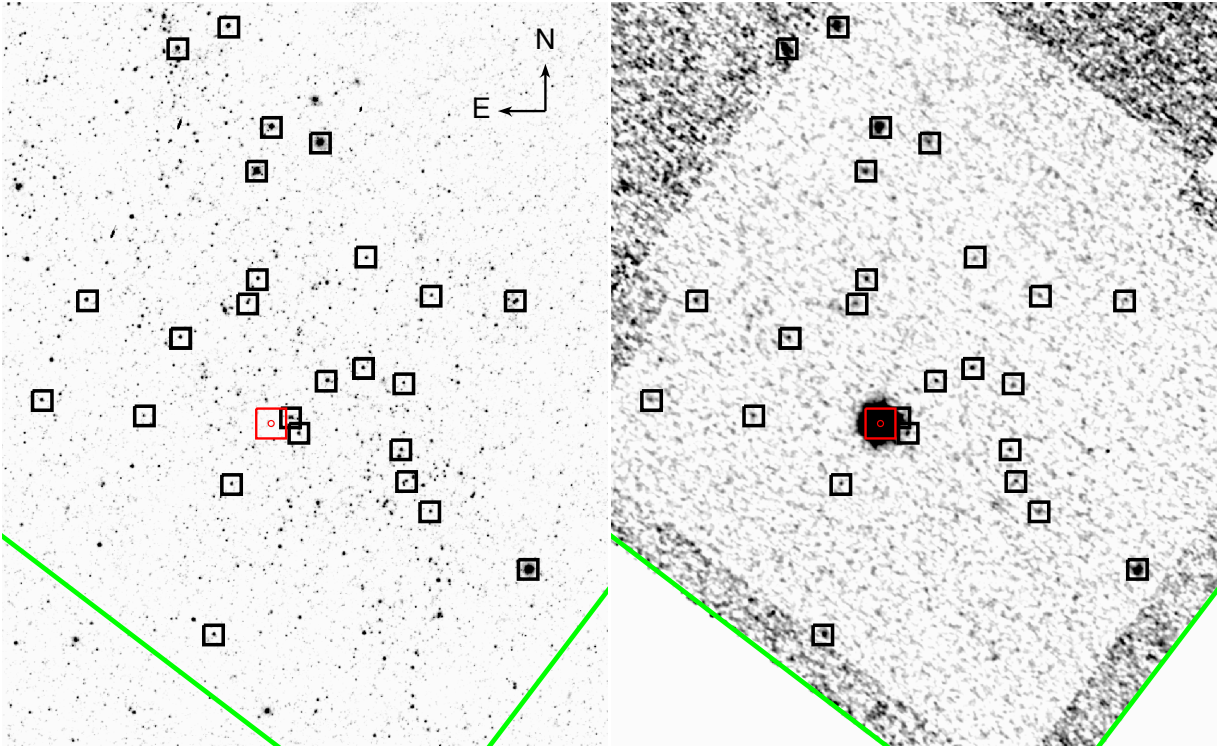

**Supplementary Figure 1** — Astrometric solution using the Keck AO image. The Keck AO image is shown in the right panel, and is astrometrically matched to a section of *HST* ACS/F814W image on the left panel. The field of view is $1' \times 1'$. Detected objects in the Keck AO images (and the counterparts in the *HST* image) are marked with black squares. 13 of them with reasonably accurate positions are used in the astrometric solution, achieving a $1\sigma$ precision of 21 mas. The red square is centred on the SN and has a scale of $2'' \times 2''$, and matches the scale of the right panel presented in Figure 1 of the main text. The green lines mark the boundary of the AO image in the ACS/F814W image.

for the sky background. Prior to registration, we applied the distortion correction provided by the Keck Observatory[3]. The dithered images were registered and stacked to produce a single image with a resolution of $0''.04$ pixel$^{-1}$, slightly superior to the resolution of the ACS images ($0''.05$ pixel$^{-1}$).

We were able to identify 13 stars around the SN position that are present in both the Keck AO and ACS/F814W images (Supplementary Figure 1). The astrometric solution of this set of objects is limited by the difficulties of accurately measuring the centroid positions for the stars on the Keck AO images due to the signal-to-noise ratios (S/N) of the detections. Nonetheless, the solution has a precision of $1\sigma = 0''.021$ or 21 mas, much improved over our reported precision (58 mas) in Li et al. (2011a)[7]. The SN location in the ACS/F814W image (filename: HST_9490_a2_ACS_WFC_F814W_sci.fits) is $X = 3309.30$, $Y = 1750.12$, with an uncertainty of 0.30 pixel in both axes. This position is consistent with our preliminary analysis[7] to within $1\sigma$ uncertainty, and also agrees with the position reported by Smartt et al. (2011)[10]. The precise position for the SN location is $\alpha = 14^h03^m05^s.733$, $\delta = +54°16'25''.18$ (J2000.0) in the world coordinate system (WCS) of the ACS/F814W image.

The site of PTF11kly in the other *HST* data reported in Supplementary Table 2 was carefully inspected, and no object was detected in any of the images. In the next section, we derive the limiting magnitudes of the *HST* data.

---

[3]See http://www2.keck.hawaii.edu/inst/nirc2/forReDoc/post_observing/dewarp.

3

| Label | 3.0σ | 2.5σ | 2.0σ |
|---|---|---|---|
| ACS/F435W | 27.42 | 27.62 | 27.87 |
| ACS/F555W | 27.04 | 27.25 | 27.49 |
| ACS/F814W | 26.35 | 26.55 | 26.81 |
| WFPC2/F606W | 26.60 | 26.80 | 27.05 |

**Supplementary Table 3** — Limiting magnitudes of the *HST* images in the Vega magnitude system.

**Limiting magnitudes**

We employed several methods to derive limiting magnitudes for the *HST* images. For method I, the photometry for many stars near the SN position was measured, and the measured magnitude that gives an error of 0.2 mag is considered to be the 5σ upper limit. The 3σ upper limit is then 0.4 mag deeper. To perform photometry of the stars, we have followed the recipes described by Sirianni et al. 2005[11] for ACS and by Holtzman et al. (1995a, 1995b)[12,13] for WFPC2. For method II, we measured the variance of the sky background in the immediate vicinity of the SN. This variance, along with the instrument characteristics, was used to calculate the brightness for objects with different S/N. For method III, we made use of the Exposure Time Calculator (ETC) available from STScI. Assuming a flat continuum in $F_\lambda$, we derived the magnitude for an object that would be detected at different S/N with the setup of the *HST* observations. The three methods yield limiting magnitudes that are in general within 0.1 mag of each other, and their average values are reported in Supplementary Table 3.

While the limiting magnitudes of the individual *HST* datasets already reach great depth, the ensemble of all the *HST* data, when considered together, offers even greater depth. We focus this analysis on the *V* band. Since the *HST* data were observed with different filters, they can only be combined if the spectral type of the object of interest is known. We have made use of the spectrum database for the different spectral types published by Castelli & Kurucz (2004)[14], augmented with blackbody spectra for extremely hot stars, and Pickles (1998)[15] spectra of very cool stars and brown dwarfs.

For each spectral type, we first derived the tightest limiting magnitude for the *V* band from the 4 individual *HST* images. Using the "synphot" package in IRAF/STSDAS, we converted the limiting magnitudes derived for the ACS/F435W (*B* band) and ACS/F814W (*I* band) into effective limiting magnitudes in the *V* band. The deepest *V*-band magnitude among the 4 measurements is then chosen as our "1-frame" limiting magnitude in the *V* band. Not surprisingly, when the spectral type is blue (such as O and B stars), the *HST* data observed with the blue filter (ACS/F435W) provide the deepest *V*-band limit, while for the red spectral types (K and M stars), the *HST* data in ACS/F814W are the deepest.

We then derived the limiting magnitude for the stacked *HST* data. We employed two methods to do this. For the first method ("time-stacked" method, hereafter), we converted the *HST* data observed with ACS/F435W, ACS/F814W, and WFPC2/F606W into "effective" ACS/F555W observations. We discuss ACS/F435W as an example. For a given spectral type, the *B*-band limiting magnitude was converted to the *V* band using synphot. Using ETC, we derived the "effective" exposure time at which ACS/F555W would reach the same *V*-band limiting magnitude (for the said spectral type). The stacked *HST* data were then considered as an ACS/F555W observation with an exposure time that is the sum of all the "effective" exposure times, and the limiting magnitudes were derived using ETC. In general, the limiting magnitude for the stacked image is deeper than the "1-frame" limits by 0.2–0.8 mag.

We also used an alternate method ("flux-stacked" method, hereafter) to determine an effective limiting *V* magnitude from "stacking" the *HST* images. We began with the same synphot calculated colours for the full range of spectral types used above. We then calculated the $\chi^2$ goodness-of-fit statistic for each spectral type, varying the *V* magnitude over a wide range, and comparing the model flux with the measurements. The calculation was performed in flux units, with the uncertainty in the measured flux derived from observed 3σ upper limits given in Supplementary Table 3. We then tabulated 2σ, 2.5σ, and 3σ effective limits to the *V* magnitude by determining where the fit probability reaches 0.9545, 0.9876, and 0.9973, respectively. The method is illustrated in Supplementary Figure 2, which shows the calculation for a K0 III model. Our results with this method were similar to those using the effective exposure time method above, and so we have averaged the two in determining our final constraints.

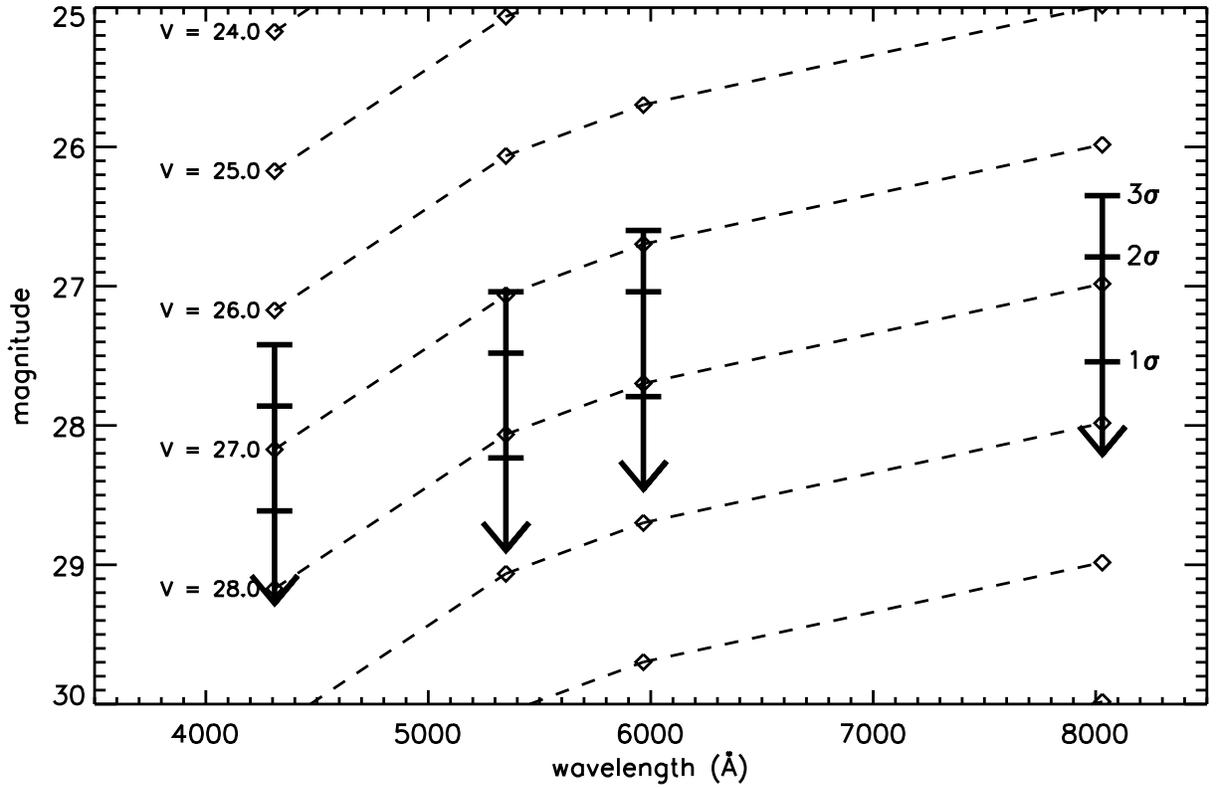

**Supplementary Figure 2** — Illustration of the $\chi^2$ method to determine the effective limiting $V$ magnitude from "stacking" the *HST* observations. The thick arrows show the observed upper limits in ACS/F435W, ACS/F555W, WFPC2/F606W, and ACS/F814W from left to right (all in the VEGAMAG system), and are hatched at $3\sigma$, $2\sigma$, and $1\sigma$ (in flux units) from top to bottom. The open diamonds (connected by dashed lines) show the synphot-calculated magnitudes for a K0 III star over a range of $V$ brightnesses. We calculated the $\chi^2$ statistic as a function of $V$ magnitude to determine the effective $V$ magnitude limits from all the data combined. In this example we find $V > 28.19$, $27.95$, and $27.75$ mag at $2\sigma$, $2.5\sigma$, and $3\sigma$, respectively.

|  | 3.0 σ | | 2.5 σ | | 2.0 σ | |
| --- | --- | --- | --- | --- | --- | --- |
| Sp. Type | "1"-frame | stacked | "1"-frame | stacked | "1"-frame | stacked |
| O3 V | 27.74 | 28.02 | 27.94 | 28.22 | 28.19 | 28.47 |
| O5 V | 27.74 | 28.03 | 27.94 | 28.23 | 28.19 | 28.48 |
| B0 V | 27.70 | 28.00 | 27.90 | 28.20 | 28.15 | 28.45 |
| B5 V | 27.58 | 27.93 | 27.78 | 28.13 | 28.03 | 28.38 |
| A0 V | 27.44 | 27.86 | 27.64 | 28.06 | 27.89 | 28.31 |
| A5 V | 27.31 | 27.82 | 27.51 | 28.02 | 27.76 | 28.27 |
| F0 V | 27.15 | 27.78 | 27.35 | 27.98 | 27.60 | 28.23 |
| F5 V | 27.10 | 27.78 | 27.30 | 27.98 | 27.55 | 28.23 |
| G0 V | 27.10 | 27.78 | 27.30 | 27.98 | 27.55 | 28.23 |
| G5 V | 27.10 | 27.78 | 27.30 | 27.98 | 27.55 | 28.23 |
| K0 V | 27.21 | 27.81 | 27.41 | 28.01 | 27.66 | 28.26 |
| K5 V | 27.77 | 28.05 | 27.97 | 28.25 | 28.22 | 28.50 |
| M0 V | 28.31 | 28.41 | 28.51 | 28.61 | 28.76 | 28.86 |
| M4 V[a] | – | 29.25 | – | 29.45 | – | 29.70 |
| M5 V[a] | – | 29.61 | – | 29.81 | – | 30.06 |
| B0 III | 27.69 | 28.00 | 27.89 | 28.20 | 28.14 | 28.45 |
| B5 III | 27.59 | 27.93 | 27.79 | 28.13 | 28.04 | 28.38 |
| G0 III | 27.10 | 27.78 | 27.30 | 27.98 | 27.55 | 28.23 |
| G5 III | 27.20 | 27.80 | 27.40 | 28.00 | 27.65 | 28.25 |
| K0 III | 27.39 | 27.86 | 27.59 | 28.06 | 27.84 | 28.31 |
| K5 III | 27.89 | 28.12 | 28.09 | 28.32 | 28.34 | 28.57 |
| M0 III | 28.25 | 28.36 | 28.45 | 28.56 | 28.70 | 28.81 |
| M5 III[a] | – | 29.28 | – | 29.48 | – | 29.73 |
| M10 III[a] | – | 30.89 | – | 31.09 | – | 31.34 |
| B0 I | 27.67 | 27.98 | 27.87 | 28.18 | 28.12 | 28.43 |
| B5 I | 27.57 | 27.93 | 27.77 | 28.13 | 28.02 | 28.38 |
| A0 I | 27.48 | 27.89 | 27.68 | 28.09 | 27.93 | 28.34 |
| A5 I | 27.44 | 27.87 | 27.64 | 28.07 | 27.89 | 28.32 |
| F0 I | 27.36 | 27.84 | 27.56 | 28.04 | 27.81 | 28.29 |
| F5 I | 27.23 | 27.81 | 27.43 | 28.01 | 27.68 | 28.26 |
| G0 I | 27.11 | 27.78 | 27.31 | 27.98 | 27.56 | 28.23 |
| G5 I | 27.40 | 27.85 | 27.60 | 28.05 | 27.85 | 28.30 |
| K0 I | 27.51 | 27.90 | 27.71 | 28.10 | 27.96 | 28.35 |
| K5 I | 28.20 | 28.32 | 28.40 | 28.52 | 28.65 | 28.77 |
| M0 I | 28.19 | 28.32 | 28.39 | 28.52 | 28.64 | 28.77 |
| M2 I | 28.77 | 28.78 | 28.97 | 28.98 | 29.22 | 29.23 |
| BB1[b] | – | 27.91 | – | 28.11 | – | 28.36 |
| BB2[b] | – | 27.94 | – | 28.14 | – | 28.39 |
| BB3[b] | – | 27.96 | – | 28.16 | – | 28.41 |

**Supplementary Table 4** — Limiting magnitudes in the $V$ band for the "stacked" $HST$ image. [a]For these spectral types, the spectra from Pickles (1998)[15] were used, and only the "flux-stacked" limits were derived. [b]Blackbody spectra with $T_{\rm eff}$ = 35,000 K (BB1), 60,000 K (BB2), and 100,000 K (BB3). For these spectra, only the "flux-stacked" limits were derived.

The reddening toward SN 2011fe is negligible in the Galaxy [$E(B-V) = 0.009$ mag][16] and in M101 (Nugent et al. 2011b). Adopting a distance modulus of $\mu = 29.05 \pm 0.23$ mag[17,18], the $V$-band magnitude limits in Supplementary Table 3 can be converted to absolute magnitudes $M(V)$. We fit a cubic spline function to the $2\sigma$ limiting magnitudes as a function of the effective temperatures of the different spectral types, and present the results in a Hertzsprung-Russell diagram (H-R diagram, Figure 2 of the main text).

For a given effective temperature and $M_V$ limit the minimum effective radius $R_{\text{eff,min}}$ of the allowed progenitor can be tabulated assuming a "bolometric correction" from the $V$-band to the total bolometric luminosity $[BC_V(T_{\text{eff}})]$[19]. At $T_{\text{eff}} = 3000$ K, $BC_V = -5.00$ mag; at $T_{\text{eff}} = 4000$ K, $BC_V = -1.12$ mag. We take $R_{\text{eff,min}}$ to be

$$R_{\text{eff,min}}/R_\odot = \frac{10^{8.4694 - 0.2[BC_V(T_{\text{eff}}) + M_V]}}{T_{\text{eff}}^2}$$

where $M_V$ is the limit we calculate above. At $T_{\text{eff}} = 3010, 3490, 4050$ K we find the limit on $M_V = 0.59, 0.24, -0.22$ mag and $R_{\text{eff,min}} = 240, 63, 32 R_\odot$ (with $R_\odot$ the mass of the Sun).

**Implications on the progenitor system of SN 2011fe from the H-R diagram**

Type Ia supernovae (SNe Ia) are believed to be caused by the thermonuclear explosion of a white dwarf composed mainly of carbon and oxygen (i.e., a CO white dwarf). In the most popular models, this is due to the degenerate ignition of carbon when the mass of the CO white-dwarf approaches the Chandrasekhar mass ($\sim 1.4 M_\odot$), the maximum possible mass.[4]

One of the main unsolved questions is what makes a white dwarf grow to the Chandrasekhar mass. There are numerous progenitor models which broadly follow into two classes: *double-degenerate (DD) models*[22,23] and *single-degenerate (SD) models*[24,25]. In the double-degenerate models, two CO white dwarfs in a close binary are driven together by the emission of gravitational waves until the lighter white dwarf is disrupted dynamically, leading to the merger of the two. A SN Ia may occur when the combined mass approaches the Chandrasekhar mass. In the single-degenerate models, the mass of the CO white dwarfs grows by accretion of matter from a companion star. Different SD models can be distinguished based on the nature of the mass donor: this can be either a main-sequence star or subgiant ("super-soft channel"[26]), a giant star ("symbiotic channel"[27]), a helium star ("helium-star channel"[24,28]), or perhaps even a degenerate object.

All of these proposed models have their advantages and problems, and there are numerous observational constraints ruling out particular models in individual cases (see, e.g., Podsiadlowski et al. 2008[29] for further discussion). In addition, it is quite possible, perhaps even likely, that more than one progenitor channel is required to explain the observed diversity among SNe Ia.

Because it is relatively nearby, SN PTF11kly provides a unique opportunity to constrain the various proposed progenitor models, at least for this SN. Figure 2 in the main text shows the upper $M_V$ limits in an H-R diagram from deep pre-SN *HST* images. The figure also indicates the location of various observed systems in our Galaxy that have been suggested to be prototypical SN Ia progenitors systems in the SD channel: RS Oph and T CrB are symbiotic binaries with orbital periods of 455 and 228 days, respectively, representing the symbiotic channel[30]; U Sco is a supersoft source with an orbital period of 30 hr[31]; and V445 Pup is a binary with a helium-star donor which underwent a helium nova in 2000 (its orbital period has not yet been determined)[32]. In all of these systems, the white dwarfs are believed to have a mass close to the Chandrasekhar mass, making them excellent candidates for SN Ia progenitors.[5]

As Figure 2 shows, the *HST* images already constrain possible progenitor models. In particular, it strongly argues against a symbiotic binary progenitor, ruling out a system like RS Oph and probably T CrB. Similarly, a helium-star binary such as V445 Pup is only marginally consistent with the upper limits, although it cannot be completely ruled out (see the blue-shaded area labelled "He-star channel" in Figure 1). Among the SD models, only the supersoft channel

---

[4]There are also alternative models in which the exploding white dwarf is below the Chandrasekhar mass, so-called sub-Chandrasekhar models. In these models, carbon may be ignited by a detonation wave caused by the detonation of helium in a shell surrounding the CO core (i.e., "double-detonation models"[20,21]).

[5]As a caveat, it should be noted that it is not established in any of these systems whether the white dwarf is a CO white dwarf or an ONeMg white dwarf. In the latter case, the white dwarf would not undergo a thermonuclear explosion when it reaches the Chandrasekhar mass, but rather experience collapse to a neutron star, not producing a SN Ia.

can easily be reconciled with the *HST* constraints. Finally, all DD models are consistent with the non-detection of a source at the position of PTF11kly, as it does not predict a bright source in the visible range.

RS Oph, T CrB, and U Sco are recurrent novae. Indeed, the recurrence time of WD binaries where the white dwarf is close to the Chandrasekhar mass is expected to be 10–20 yr[33]. As Supplementary Figure 3 shows, no nova has been detected for the last 12 yr. However, this does not rule out this channel, as the recurrence time just before the SN is quite uncertain and recurrent novae are only very bright for a very short period that could easily have been missed. Indeed, as the analysis in the following section shows, there is only a 63% probability that we would catch a nova outburst had it occurred randomly in the 6 yr preceding the SN explosion.

In summary, the progenitor models that are most consistent with the historical constraint are either the supersoft SD channel or any DD model.

## Notes on the constraints from observed SN Ia progenitor candidates

Figure 2 in the main text indicates the location in the H-R diagram of observed SN Ia candidates. For these systems we plot the effective temperature based on estimates of the spectral type (if available) or theoretical modelling of the donor star. The absolute magnitudes are taken from observations of these systems in quiescence and include, in addition to the light from the donor star, the contributions from the hot accreting white dwarf and the accretion disk, which may be quite substantial. The large uncertainties in the inferred absolute magnitudes are largely due to uncertainties in the distance estimates and the applicable extinctions.

*RS Oph and T CrB*

RS Oph and T CrB are symbiotic binaries where the white dwarf is close to the Chandrasekhar mass and experiences recurrent novae. RS Oph during quiescence has $m_V$ varying between $\sim 11$ and $\sim 11.5$ mag[34]. Recent distance estimates range from 1.4 kpc (from modelling of the last outburst) to 3.1 kpc (assuming the secondary is filling its Roche lobe; see Barry et al. 2008[35] for discussion). Using an estimate of the visual extinction of 2 mag (Iijima 2008)[36] then leads to an estimated range of $M_V$ from $-1.2$ to $-3.5$. The spectral type of the donor has been determined to be M0 to M2 III (Anupama & Mikołajewska 1999)[37], implying an effective temperature of 3750 to 3900 K (Straizys & Kuriliene 1981)[38].

RS Oph has a quiescent $m_V \approx 10$ mag. With a distance estimate in the range of 0.9 to 1.3 kpc and an adopted extinction of $\sim 0.5$ mag (Belczynski & Mikołajewska 1998)[39], this leads to an absolute $M_V$ of $-1$ to $+0$ mag. Its estimated spectral type of M3/4 III implies an effective temperature for the donor of 3550 to 3650 K (Straizys & Kuriliene 1981)[38].

*U Sco*

U Sco is supersoft recurrent nova with a very massive white dwarf. Since its distance is not well established, we plot typical parameters for a system of this type from the modelling by Podsiadlowski (2003)[40].[6] Adopting a reasonable range in properties leads to a range in $M_V$ 0f 2.5 to 3.9 mag and a range of $T_{\rm eff}$ of 5200 to 6000 K.

*V 445 Pup*

V 445 Pup experienced a helium nova in 2000 and is a prototypical system for a SD model with a helium-star donor. Following Kato et al. (2008)[32], we adopt a quiescent $M_V$ of 14.5 mag. Its distance is uncertain but has been constrained to be between 3.5 and 6.5 kpc, and its $A_V$ has to be estimated to be $\sim 1.6$ mag (Kato et al. 2008). Taking a range of 1 to 2 mag for the latter leads to a range in the absolute magnitude of V445 Pup during quiescence of $-1.5$ to $+1$ mag. The effective temperature of the donor star is not well determined, but helium stellar evolution tracks (Fig. 6 in Kato et al. 2008) suggest a temperature betwen 25,000 K and 60,000 K.

*He-star channel*

For the He-star channel we take the theoretical models of Liu et al. (2010)[28], taking the properties of the donor star at the time of explosion (their Fig. 6) and applying appropriate bolometric corrections from Torres (2010)[19].

---

[6]These parameters are typical for progenitors in the supersoft channel, regardless of whether the white dwarf in U Sco is a CO or ONeMg white dwarf.

8

## Historical light curves

Regular ground-based optical imaging of M101 has been obtained with four facilities during the course of the past ~13 years: the 0.76-m Katzman Automatic Imaging Telescope (KAIT) on Mount Hamilton, CA, as part of the Lick Observatory Supernova Search (LOSS;[41]); the Palomar 60-in telescope (P60;[42]) as part of the P60 Fast Transients in Nearest Galaxies survey (FasTING;[43]); the 3.6-m Canada-France-Hawaii Telescope on Mauna Kea; and the Palomar 48-in as part of the ongoing Palomar Transient Factory[44,45]. In addition to the ground-based imaging, M101 was observed multiple times by *HST*; see the discussion in the previous section of the supplementary material. All of the ground-based images were reduced using standard techniques, including removal of the bias and dark current from each frame followed by proper flatfielding.

The projected position of PTF11kly/SN 2011fe lies very close to a spiral arm structure within M101, leading to a complex background region in the vicinity of the SN (see Figure 1). Thus, in order to search for any variable optical emission that may be associated with the progenitor system of PTF11kly, we employed image subtraction to remove the complex, but presumably static, background signal. For each individual image (hereafter, new image), a reference image was constructed using a deep coaddition of several images taken with the same camera and in the same filter but during a different year than the new image to ensure that any possible variable emission in the new frames is not also present in the reference image. Following the construction of appropriate reference images, for each new image the corresponding reference was then astrometrically aligned with the new image, both images were convolved to have the same seeing, and the flux in the reference image was then subtracted from the flux in the new image.

Following proper subtraction, we examined whether any variable emission can be detected at the location of the SN. We measured the flux in an aperture centred on the SN position, which is precisely measured to an accuracy < 1 pixel by transforming a P48 image that contains the SN to the same coordinates as the subtracted image. To determine the uncertainty in this single-aperture measurement, we measured the variance in 1000 apertures randomly placed within a $\sim 90'' \times 90''$ box centred on the SN position, which ensures similar background properties as the SN position but excludes strong systematic variations related to residuals associated with poorly subtracted bright stars. Using this method there are multiple images where we detect flux at the SN position with greater than $3\sigma$ confidence; however, since we have $\sim 3000$ images of M101, we would expect about five $3\sigma$ detections based on Gaussian fluctuations. In order to test the null hypothesis that no emission is present in the subtraction images, with the alternative hypothesis that emission is present in an image, we employ the false-discovery rate method (FDR; see[46,47]) which controls for the fact that in multiple hypothesis testing, in this case many images, there are multiple comparisons to the null hypothesis. Using the FDR method, and setting $\alpha = 0.05$ (equivalent to a $2\sigma$ detection), we are unable to reject the null hypothesis that no variable emission is present at the position of the SN. This result is not surprising given that the typical individual images from KAIT ($m_{\rm clear}(3\sigma) \approx 18.8$ mag), P48 ($m_{\rm Mould-R}(3\sigma) \approx 21$ mag), and P60 ($m_g(3\sigma) \approx 21.5$ mag) are too shallow to detect all but the very brightest novae (see Figure 3), and even that would require an image to be taken within a few days of the peak brightness of the nova. While the CFHT images are deep enough to detect even the very faintest novae, since only 7 images were taken over slightly more than 7 yr the probability of a nova being bright enough for detection concurrently with an acquired image is only $\sim 7\%$.

To further constrain the possibility of pre-SN variability, we stack individual images grouped closely together in time in order to obtain limits that are deeper than those in the individual frames. Should variable emission, such as that associated with a nova, be present within the data, then the stacked images will reflect the average observed flux from each of the individual images. On week-to-month timescales the detection threshold for stacked images decreases *faster* than the average flux from a nova, and therefore stacked images will allow an improved opportunity for the detection of variability on timescales similar to those of a nova. For KAIT we stacked images on intervals of 30 days and for entire observing seasons. These images extend over the full lifetime of LOSS, $\sim 13$ yr, and are shown in black in Figure 3. For the P60 we employ both 7 day and 30 day stacks which extend from late 2006 through mid 2009, shown in green in Figure 3. For the P48 observations of M101 we employ 7 day and 30 day stacks extending over the full lifetime of PTF from mid-2009 to the present. These limits are shown in red in Figure 3. Following the production of stacked images, we use the same procedure described above to test for variable emission: we subtract the flux from a template image and measure the flux at the SN position compared with several randomly placed apertures in the vicinity of the SN position. Using the FDR method, we once again find no significant detection of variable emission at the position of the SN.

Following the non-detection of any pre-SN variability, we attempt to constrain the probability that a nova could

9

have occurred in the 13 yr prior to PTF11kly/SN 2011fe but that gaps in our observations or insufficient depth in our stacked images prevented us from detecting the signal. The true characterisation of this probability requires an *a priori* knowledge of the nova luminosity function (NLF) including the distribution of nova rise times, which are typically short (see, e.g.,[43]). To the best of our knowledge, a complete sample of the distribution of the NLF has not been assembled to date, and thus we must use proxies in our estimates of the probability of detecting a nova signal. As a proxy for the true NLF we use the light curves of two novae discovered during the FasTING survey in M31, nova 2008-09a and nova 2008-09c[43]. We elect to use extragalactic novae to avoid the large systematic uncertainties associated with measuring the distance and reddening toward Galactic novae, and in particular, the two selective novae have strong constraints on the rise time as a result of the high-cadence observations from FasTING. We can characterise these two novae using the maximum-magnitude vs. rate-of-decline (MMRD) relation (for a theoretical explanation see, e.g.,[48], although see also[43]), from which we learn that nova 2008-09c is a typical nova with $M_g = -7.8$ mag and a time to decline from peak by 1 mag, $t_1 = 9.1$ days[43]. Nova 2008-09a, on the other hand, is both faint, $M_g = -6.8$ mag, and fast evolving, $t_1 = 6.3$ days[43], meaning it is among the most challenging type of nova to detect. Using a smooth spline fit to the data for these two novae, we deredden them and shift them to the distance of M101 to serve as a proxy to the true NLF. The light curves for each of these novae are shown in Figure 3: 2008-09c is in green, while 2008-09a is in blue.

The typical detection limits from KAIT are too shallow to detect all but the most luminous novae, and even in these cases the signal would often be degraded given the large number of KAIT images needed to reach a detection threshold of $m > 21.5$ mag. Therefore, we focus our efforts on constraining the presence of a nova during the time of the P60 and PTF observations, which constrains the following limits to the past five observing seasons starting from late 2006 until the time of the SN (see the bottom panel of Figure 3). To determine the detection probabilities, at random times we insert a fake nova signal into the data and determine the average flux from the nova based on the actual observation times of the stacked images that overlap the nova in time. If the average flux is greater than the measured detection threshold from the stacked image, then the nova is considered detected. We have repeated this procedure 1 million times for each of our example novae, and we find a detection probability of 2008-09a-like and 2008-09c-like novae of 0.440 and 0.628, respectively.

It is important to note, however, that these detection limits provide a *lower limit* to the true detection probability since we have adopted a stacking scheme that does not include overlapping windows (the precise scheme we use can be visualised using the horizontal error bars of the limits shown in Figure 3). Under this method a nova that peaks directly in between two stacking windows could escape detection, as each window overlaps with only a fraction of the nova. To prevent this from happening, an additional window, preferably centred near the time of peak brightness from the nova, should be created. Given the broad complexities associated with the image sequences, for example variable cadence and detection thresholds, the ideal method to test for the presence of a nova would be to stack all the images in every possible grouping on timescales shorter than those typical for a nova, $\sim 30$ days. Such an effort would be impractical for the $\sim 3000$ pre-SN images of M101 discussed here, and is beyond the scope of the current paper.

## X-ray upper limits

We analysed 11 epochs of *Chandra* observations of the field containing PTF11kly, all taken in 2004. Additional observations in which the SN location is near one of the CCD edges were removed. We detected no bright source in any epochs, and we additionally sub-selected the epochs capable of producing the most stringent flux limit. These were the 4 epochs (ObsIDs 6114, 6115, 4735, and 6118) where the SN was within $3.5'$ of the observation aimpoint, placing the source on the most sensitive ACIS-S3 detector. The total exposure in these 4 epochs is 142 ks.

Count rates and spectra were extracted from a 4-pixel ($2''$) radius aperture, containing 90% of the encircled photon flux, around the SN position for each epoch using standard tools in CIAO 4.3[7]. The sky background was estimated using an annular extraction region with an inner radius of 20 pixels and an outer radius of 50 pixels about the source position. Given 9 counts (0.3–8 keV) detected in the source region over an expected background of 11.2 counts, we determine a $3\sigma$ upper limit on the source count rate of $9.1 \times 10^{-5}$ count s$^{-1}$.

We fit the summed spectra in ISIS[8] by minimizing the Cash (1979) statistic[49]. Assuming an absorbed power-law

---

[7]http://cxc.harvard.edu/ciao/index.html.
[8]http://space.mit.edu/CXC/ISIS.

10

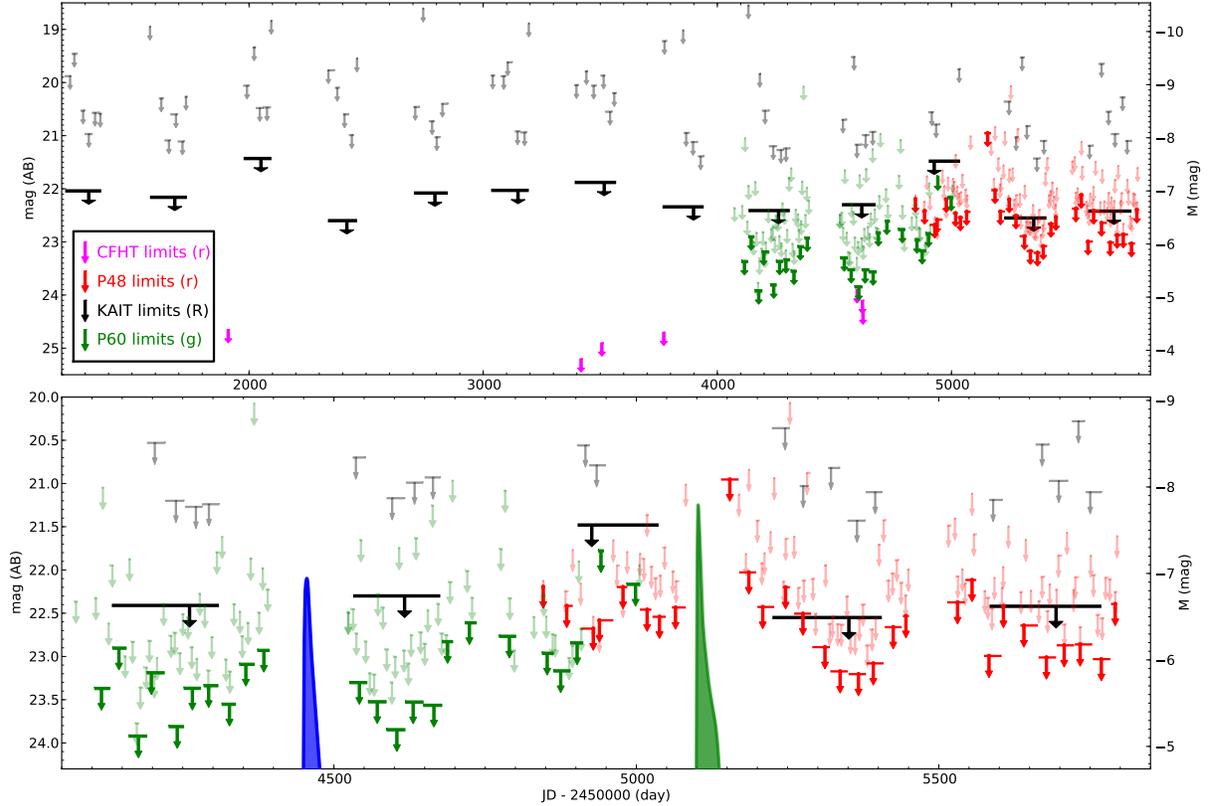

**Supplementary Figure 3 — Historical limits on pre-SN emission from the progenitor system of PTF11kly/SN 2011fe.** Shown in the top panel are the limiting magnitudes for stacked images from pre-SN imaging of M101 from KAIT (R band), CFHT (R/r band), P60 (g band), and PTF ($R_{\rm Mould}$) extending > 12 yr prior to the SN explosion. The KAIT data are stacked on both monthly (light black) and yearly (dark black) intervals, CHFT limits are shown in magenta, P60 data are stacked on both weekly (light green) and monthly (dark green) intervals, and the PTF/P48 data are stacked on weekly (light red) and monthly (dark red) intervals. The width of the error bars indicates the time elapsed between the first image and the last image included in a stack. The bottom panel shows a zoom in on the ∼ 5 yr prior to the SN, when the P60 and P48 data provide the strongest constraints on a possible nova. Overplotted on the data are the light curves of two novae from M31: 2008-09a, a quickly-fading, faint nova (blue), and 2008-09c, a typical classical nova (green;[43]), with their apparent brightness scaled to how they would appear in M101. The limits show we would have been capable of detecting such outbursts, with the probability of detection ∼ 44% and ∼ 63% for 2008-09a-like novae and 2008-09c-like novae, respectively.

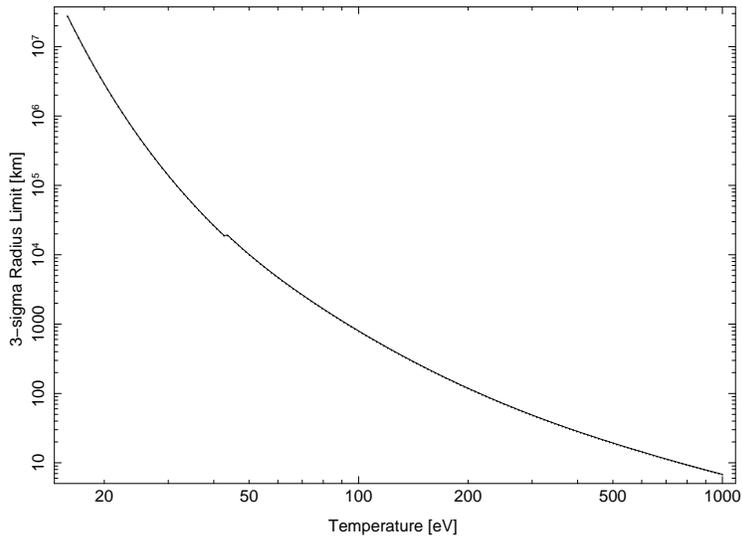

**Supplementary Figure 4 —** The limit on the progenitor radius as a function of temperature assuming a blackbody model, derived from the X-ray luminosity limits.

model with photon index $\Gamma = 2$ and Galactic column density $N_H = 1.8 \times 10^{20}$ cm$^{-2}$ [50], we determine a $3\sigma$ limiting unabsorbed flux of $4.1 \times 10^{36}$ erg s$^{-1}$ (0.3–8 keV). For a 67 eV blackbody (e.g., Gilfanov & Bogdan, arXiv 1103.3659; see, however, Izumi et al. 2010[51], the bolometric flux limit is $2.5 \times 10^{37}$ erg s$^{-1}$. Supplementary Figure 4 displays the limit on the progenitor radius as a function of temperature, assuming a blackbody model.

## Spitzer data reduction

We analysed pre-explosion *Spitzer*/IRAC data taken on UT 2004 March 8.09 and 2004 March 8.49 (PID:60, PI: G. Reike). Using 12 stars on the PTF discovery images[5] that are also present in the *Spitzer* images, we were able to locate the SN position with a $1\sigma$ uncertainty of $0''.132$ (132 mas). No point source was detected in any of the *Spitzer* data. Limiting magnitudes were derived in a $20 \times 20$ pixel box centred on the location of PTF11kly; see Supplementary Table 4.

According to a mid-infrared (IR) H-R diagram presented by Ardila et al. (2010)[52], the mid-IR luminosities for red giants (K0 III to M6 III) are in the range of $4 \times 10^{35}$ to $4 \times 10^{36}$ erg s$^{-1}$. Thus, the limits from *Spitzer* data alone rule out the brightest red giants, but not red giants with moderate to low luminosities.

The *Spitzer* mid-IR upper limits are potentially useful as constraints on the DD models. Shen et al. (2011)[53] suggests that, for a DD merger to produce a SN Ia, it needs to evolve and cool slowly, for $> 1 \times 10^6$ yr. This means the merger product would radiate at smaller than a fraction ($\sim 1$ %) of the Eddington limit ($\sim 4 \times 10^{38}$ erg s$^{-1}$), i.e., $< \sim 4 \times 10^{36}$ erg s$^{-1}$. The nondetections in the *Spitzer* data are thus consistent with the requirement of low IR luminosity from the DD models.

| MJD | Filter ($\mu$m) | Absolute Mag Limit (Vega) | $\nu F_\nu$ Limit ($10^{36}$ erg s$^{-1}$) |
|---|---|---|---|
| 53072.090 | 3.6 | $-10.6$ | 1.4 |
| 53072.490 | 3.6 | $-10.6$ | 1.2 |
| 53072.090 | 4.5 | $-11.2$ | 1.2 |
| 53072.490 | 4.5 | $-11.1$ | 1.0 |
| 53072.090 | 5.8 | $-13.1$ | 11 |
| 53072.490 | 5.8 | $-13.1$ | 13 |
| 53072.090 | 8.0 | $-13.8$ | 6.7 |
| 53072.490 | 8.0 | $-13.9$ | 8.6 |

**Supplementary Table 5** — *Spitzer* 3$\sigma$ upper limits.

# References

1. Smartt, S. Limits on Ia progenitors from pre-explosion imaging: Lessons from core-collapse progenitor studies. Observational signatures of Type Ia Supernova Progenitors. Http://www.lorentzcenter.nl/lc/web/2010/391/presentations/Smartt.ppt.

2. Maoz, D. & Mannucci, F. A search for the progenitors of two Type Ia Supernovae in NGC 1316. MNRAS **388**, 421–428 (2008).

3. Nelemans, G., Voss, R., Roelofs, G. & Bassa, C. Limits on the X-ray and optical luminosity of the progenitor of the Type Ia supernova 2007sr. MNRAS **388**, 487–494 (2008).

4. Voss, R. & Nelemans, G. Discovery of the progenitor of the type Ia supernova 2007on. Nature **451**, 802–804 (2008).

5. Nugent, P. E. *et al.* Young Type Ia Supernova PTF11kly in M101. *The Astronomer's Telegram* **3581**, 1 (2011).

6. Li, W. *et al.* Identification of the Red Supergiant Progenitor of Supernova 2005cs: Do the Progenitors of Type II-P Supernovae Have Low Mass? ApJ **641**, 1060–1070 (2006).

7. Li, W., Nugent, P., Bloom, J. S. & Filippenko, A. V. Analysis of the archival HST images of PTF11kly in M101. *The Astronomer's Telegram* **3582**, 1 (2011).

8. Li, W., Nugent, P., Bloom, J. S., Filippenko, A. V. & Jha, S. Further Analysis of the archival HST images of PTF11kly in M101. *The Astronomer's Telegram* **3585**, 1 (2011).

9. Wizinowich, P. L. *et al.* The w. m. keck observatory laser guide star adaptive optics system: Overview. *PASP* **118**, 297 (2006).

10. Smartt, S. J., Fraser, M., Kotak, R., Magill, L. & Maguire, K. No progenitor detection for PTF11kly/SN2011fe in Hubble Space Telescope pre-explosion images. *The Astronomer's Telegram* **3623**, 1 (2011).

11. Sirianni, M. *et al.* The Photometric Performance and Calibration of the Hubble Space Telescope Advanced Camera for Surveys. PASP **117**, 1049–1112 (2005).

12. Holtzman, J. A. *et al.* The performance and calibration of WFPC2 on the Hubble Space Telescope. PASP **107**, 156–178 (1995).

13. Holtzman, J. A. *et al.* The Photometric Performance and Calibration of WFPC2. PASP **107**, 1065–+ (1995).

14. Castelli, F. & Kurucz, R. L. New Grids of ATLAS9 Model Atmospheres. *ArXiv Astrophysics e-prints* (2004). `arXiv:astro-ph/0405087`.

15. Pickles, A. J. A Stellar Spectral Flux Library: 1150-25000 Å. PASP **110**, 863–878 (1998).

16. Schlegel, D. J., Finkbeiner, D. P. & Davis, M. Maps of Dust Infrared Emission for Use in Estimation of Reddening and Cosmic Microwave Background Radiation Foregrounds. ApJ **500**, 525–553 (1998). `arXiv:astro-ph/9710327`.

17. Stetson, P. B. *et al.* The Extragalactic Distance Scale Key Project. XVI. Cepheid Variables in an Inner Field of M101. ApJ **508**, 491–517 (1998).

18. Shappee, B. J. & Stanek, K. Z. A New Cepheid Distance to the Giant Spiral M101 Based on Image Subtraction of Hubble Space Telescope/Advanced Camera for Surveys Observations. ApJ **733**, 124–148 (2011).

19. Torres, G. On the Use of Empirical Bolometric Corrections for Stars. AJ **140**, 1158–1162 (2010).

20. Woosley, S. E. & Weaver, T. A. Sub-Chandrasekhar mass models for Type Ia supernovae. ApJ **423**, 371–379 (1994).

21. Fink, M. *et al.* Double-detonation sub-Chandrasekhar supernovae: can minimum helium shell masses detonate the core? A&A **514**, A53+ (2010). `1002.2173`.

22. Iben, I., Jr. & Tutukov, A. V. Supernovae of type I as end products of the evolution of binaries with components of moderate initial mass (M not greater than about 9 solar masses). ApJS **54**, 335–372 (1984).

23. Webbink, R. F. Double white dwarfs as progenitors of R Coronae Borealis stars and Type I supernovae. ApJ **277**, 355–360 (1984).

24. Nomoto, K. Accreting white dwarf models for type I supernovae. I - Presupernova evolution and triggering mechanisms. ApJ **253**, 798–810 (1982).

25. Whelan, J. & Iben, I., Jr. Binaries and Supernovae of Type I. ApJ **186**, 1007–1014 (1973).

26. van den Heuvel, E. P. J., Bhattacharya, D., Nomoto, K. & Rappaport, S. A. Accreting white dwarf models for CAL 83, CAL 87 and other ultrasoft X-ray sources in the LMC. A&A **262**, 97–105 (1992).

27. Munari, U. & Renzini, A. Are symbiotic stars the precursors of type IA supernovae? ApJ **397**, L87–L90 (1992).

28. Liu, W.-M., Chen, W.-C., Wang, B. & Han, Z. W. Helium-star evolutionary channel to super-Chandrasekhar mass type Ia supernovae. A&A **523**, A3+ (2010).

29. Podsiadlowski, P., Mazzali, P., Lesaffre, P., Han, Z. & Förster, F. The nuclear diversity of Type Ia supernova explosions. New Ast. Rev. **52**, 381–385 (2008).

30. Hachisu, I. & Kato, M. Recurrent Novae as a Progenitor System of Type Ia Supernovae. I. RS Ophiuchi Subclass: Systems with a Red Giant Companion. ApJ **558**, 323–350 (2001).

31. Thoroughgood, T. D., Dhillon, V. S., Littlefair, S. P., Marsh, T. R. & Smith, D. A. The mass of the white dwarf in the recurrent nova U Scorpii. MNRAS **327**, 1323–1333 (2001). `arXiv:astro-ph/0107477`.

32. Kato, M., Hachisu, I., Kiyota, S. & Saio, H. Helium Nova on a Very Massive White Dwarf: A Revised Light-Curve Model of V445 Puppis (2000). ApJ **684**, 1366–1373 (2008).

33. Schaefer, B. E. Comprehensive Photometric Histories of All Known Galactic Recurrent Novae. ApJS **187**, 275–373 (2010).

34. Gromadzki, M., Mikolajewska, J. & Lachowicz, P. Post-outburst Variations in the Optical Light Curve of RS˜Ophiuchi. In A. Evans, M. F. Bode, T. J. O'Brien, & M. J. Darnley (ed.) *RS Ophiuchi (2006) and the Recurrent Nova Phenomenon*, vol. 401 of *Astronomical Society of the Pacific Conference Series*, 219–+ (2008).

35. Barry, R. K. *et al.* On the Distance of RS Ophiuchi. In A. Evans, M. F. Bode, T. J. O'Brien, & M. J. Darnley (ed.) *RS Ophiuchi (2006) and the Recurrent Nova Phenomenon*, vol. 401 of *Astronomical Society of the Pacific Conference Series*, 52–+ (2008).

36. Iijima, T. Circumstellar Envelope of RS Ophiuchi. In A. Evans, M. F. Bode, T. J. O'Brien, & M. J. Darnley (ed.) *RS Ophiuchi (2006) and the Recurrent Nova Phenomenon*, vol. 401 of *Astronomical Society of the Pacific Conference Series*, 115–+ (2008).

37. Anupama, G. C. & Mikołajewska, J. Recurrent novae at quiescence: systems with giant secondaries. A&A **344**, 177–187 (1999). `arXiv:astro-ph/9812432`.

38. Straizys, V. & Kuriliene, G. Fundamental stellar parameters derived from the evolutionary tracks. Ap&SS **80**, 353–368 (1981).

39. Belczynski, K. & Mikolajewska, J. New binary parameters for the symbiotic recurrent nova T Coronae Borealis. MNRAS **296**, 77–84 (1998). `arXiv:astro-ph/9711151`.

40. Podsiadlowski, P. On the Evolution and Appearance of a Surviving Companion after a Type Ia Supernova Explosion. *ArXiv Astrophysics e-prints* (2003). arXiv:astro-ph/0303660.

41. Filippenko, A. V., Li, W. D., Treffers, R. R. & Modjaz, M. The Lick Observatory Supernova Search with the Katzman Automatic Imaging Telescope. In B. Paczynski, W.-P. Chen, & C. Lemme (ed.) *IAU Colloq. 183: Small Telescope Astronomy on Global Scales*, vol. 246 of *Astronomical Society of the Pacific Conference Series*, 121 (2001).

42. Cenko, S. B. *et al.* The Automated Palomar 60 Inch Telescope. PASP **118**, 1396–1406 (2006).

43. Kasliwal, M. M. *et al.* Discovery of a New Photometric Sub-class of Faint and Fast Classical Novae. ApJ **735**, 94–+ (2011).

44. Rau, A. *et al.* Exploring the Optical Transient Sky with the Palomar Transient Factory. PASP **121**, 1334–1351 (2009).

45. Law, N. M. *et al.* The Palomar Transient Factory: System Overview, Performance, and First Results. PASP **121**, 1395–1408 (2009).

46. Benjamini, Y. & Hochberg, Y. Controlling the false discovery rate: A practical and powerful approach to multiple testing. *Journal of the Royal Statistical Society. Series B (Methodological)* **57**, pp. 289–300 (1995). URL http://www.jstor.org/stable/2346101.

47. Miller, C. J. *et al.* Controlling the False-Discovery Rate in Astrophysical Data Analysis. AJ **122**, 3492–3505 (2001). arXiv:astro-ph/0107034.

48. Livio, M. Classical novae and the extragalactic distance scale. ApJ **393**, 516–522 (1992).

49. Cash, W. Parameter estimation in astronomy through application of the likelihood ratio. ApJ **228**, 939–947 (1979).

50. Kalberla, P. M. W. *et al.* The Leiden/Argentine/Bonn (LAB) Survey of Galactic HI. Final data release of the combined LDS and IAR surveys with improved stray-radiation corrections. A&A **440**, 775–782 (2005). arXiv:astro-ph/0504140.

51. Hachisu, I., Kato, M. & Nomoto, K. Supersoft X-ray Phase of Single Degenerate Type Ia Supernova Progenitors in Early-type Galaxies. ApJ **724**, L212–L216 (2010). 1010.5860.

52. Ardila, D. R. *et al.* SpS1-The Spitzer atlas of stellar spectra. *Highlights of Astronomy* **15**, 512–512 (2010). 1010.5618.

53. Shen, K. J., Bildsten, L., Kasen, D. & Quataert, E. The Long-Term Evolution of Double White Dwarf Mergers. *ArXiv e-prints* (2011). 1108.4036.




\vskip 0.7in

\begin{thebibliography}{10}
\expandafter\ifx\csname url\endcsname\relax
  \def\url#1{\texttt{#1}}\fi
\expandafter\ifx\csname urlprefix\endcsname\relax\def\urlprefix{URL }\fi
\providecommand{\bibinfo}[2]{#2}
\providecommand{\eprint}[2][]{\url{#2}}

\bibitem{1993ApJ...413L.105P}
\bibinfo{author}{{Phillips}, M.~M.}
\newblock \bibinfo{title}{{The absolute magnitudes of Type IA supernovae}}.
\newblock \emph{\bibinfo{journal}{\apjl}} \textbf{\bibinfo{volume}{413}},
  \bibinfo{pages}{L105--L108} (\bibinfo{year}{1993}).

\bibitem{2011NatCo...2E.350H}
\bibinfo{author}{{Howell}, D.~A.}
\newblock \bibinfo{title}{{Type Ia supernovae as stellar endpoints and
  cosmological tools}}.
\newblock \emph{\bibinfo{journal}{Nature Communications}}
  \textbf{\bibinfo{volume}{2}} (\bibinfo{year}{2011}).
\newblock \eprint{1011.0441}.

\bibitem{1998AJ....116.1009R}
\bibinfo{author}{{Riess}, A.~G.} \emph{et~al.}
\newblock \bibinfo{title}{{Observational Evidence from Supernovae for an
  Accelerating Universe and a Cosmological Constant}}.
\newblock \emph{\bibinfo{journal}{\aj}} \textbf{\bibinfo{volume}{116}},
  \bibinfo{pages}{1009--1038} (\bibinfo{year}{1998}).

\bibitem{1999ApJ...517..565P}
\bibinfo{author}{{Perlmutter}, S.} \emph{et~al.}
\newblock \bibinfo{title}{{Measurements of Omega and Lambda from 42
  High-Redshift Supernovae}}.
\newblock \emph{\bibinfo{journal}{\apj}} \textbf{\bibinfo{volume}{517}},
  \bibinfo{pages}{565--586} (\bibinfo{year}{1999}).

\bibitem{1982ApJ...253..798N}
\bibinfo{author}{{Nomoto}, K.}
\newblock \bibinfo{title}{{Accreting white dwarf models for type I supernovae.
  I - Presupernova evolution and triggering mechanisms}}.
\newblock \emph{\bibinfo{journal}{\apj}} \textbf{\bibinfo{volume}{253}},
  \bibinfo{pages}{798--810} (\bibinfo{year}{1982}).

\bibitem{1984ApJS...54..335I}
\bibinfo{author}{{Iben}, I., Jr.} \& \bibinfo{author}{{Tutukov}, A.~V.}
\newblock \bibinfo{title}{{Supernovae of type I as end products of the
  evolution of binaries with components of moderate initial mass (M not greater
  than about 9 solar masses)}}.
\newblock \emph{\bibinfo{journal}{\apjs}} \textbf{\bibinfo{volume}{54}},
  \bibinfo{pages}{335--372} (\bibinfo{year}{1984}).

\bibitem{1998ApJ...508..491S}
\bibinfo{author}{{Stetson}, P.~B.} \emph{et~al.}
\newblock \bibinfo{title}{{The Extragalactic Distance Scale Key Project. XVI.
  Cepheid Variables in an Inner Field of M101}}.
\newblock \emph{\bibinfo{journal}{\apj}} \textbf{\bibinfo{volume}{508}},
  \bibinfo{pages}{491--517} (\bibinfo{year}{1998}).

\bibitem{2011ApJ...733..124S}
\bibinfo{author}{{Shappee}, B.~J.} \& \bibinfo{author}{{Stanek}, K.~Z.}
\newblock \bibinfo{title}{{A New Cepheid Distance to the Giant Spiral M101
  Based on Image Subtraction of Hubble Space Telescope/Advanced Camera for
  Surveys Observations}}.
\newblock \emph{\bibinfo{journal}{\apj}} \textbf{\bibinfo{volume}{733}},
  \bibinfo{pages}{124--148} (\bibinfo{year}{2011}).

\bibitem{thenuge}
\bibinfo{author}{{Nugent}, P.} \emph{et~al.}
\newblock \bibinfo{title}{{Discovery of a Type Ia Supernova Within Hours of
  Explosion in the Pinwheel Galaxy}} (\bibinfo{year}{2011}).
\newblock \bibinfo{note}{Submitted}.

\bibitem{1997ARA&A..35..309F}
\bibinfo{author}{{Filippenko}, A.~V.}
\newblock \bibinfo{title}{{Optical Spectra of Supernovae}}.
\newblock \emph{\bibinfo{journal}{\araa}} \textbf{\bibinfo{volume}{35}},
  \bibinfo{pages}{309--355} (\bibinfo{year}{1997}).

\bibitem{2007Sci...315..825M}
\bibinfo{author}{{Mazzali}, P.~A.}, \bibinfo{author}{{R{\"o}pke}, F.~K.},
  \bibinfo{author}{{Benetti}, S.} \& \bibinfo{author}{{Hillebrandt}, W.}
\newblock \bibinfo{title}{{A Common Explosion Mechanism for Type Ia
  Supernovae}}.
\newblock \emph{\bibinfo{journal}{Science}} \textbf{\bibinfo{volume}{315}},
  \bibinfo{pages}{825--828} (\bibinfo{year}{2007}).

\bibitem{1997Sci...276.1378N}
\bibinfo{author}{{Nomoto}, K.}, \bibinfo{author}{{Iwamoto}, K.} \&
  \bibinfo{author}{{Kishimoto}, N.}
\newblock \bibinfo{title}{{Type Ia supernovae: their origin and possible
  applications in cosmology.}}
\newblock \emph{\bibinfo{journal}{Science}} \textbf{\bibinfo{volume}{276}},
  \bibinfo{pages}{1378--1382} (\bibinfo{year}{1997}).

\bibitem{2008NewAR..52..381P}
\bibinfo{author}{{Podsiadlowski}, P.}, \bibinfo{author}{{Mazzali}, P.},
  \bibinfo{author}{{Lesaffre}, P.}, \bibinfo{author}{{Han}, Z.} \&
  \bibinfo{author}{{F{\"o}rster}, F.}
\newblock \bibinfo{title}{{The nuclear diversity of Type Ia supernova
  explosions}}.
\newblock \emph{\bibinfo{journal}{\nar}} \textbf{\bibinfo{volume}{52}},
  \bibinfo{pages}{381--385} (\bibinfo{year}{2008}).

\bibitem{1984ApJ...277..355W}
\bibinfo{author}{{Webbink}, R.~F.}
\newblock \bibinfo{title}{{Double white dwarfs as progenitors of R Coronae
  Borealis stars and Type I supernovae}}.
\newblock \emph{\bibinfo{journal}{\apj}} \textbf{\bibinfo{volume}{277}},
  \bibinfo{pages}{355--360} (\bibinfo{year}{1984}).

\bibitem{1973ApJ...186.1007W}
\bibinfo{author}{{Whelan}, J.} \& \bibinfo{author}{{Iben}, I., Jr.}
\newblock \bibinfo{title}{{Binaries and Supernovae of Type I}}.
\newblock \emph{\bibinfo{journal}{\apj}} \textbf{\bibinfo{volume}{186}},
  \bibinfo{pages}{1007--1014} (\bibinfo{year}{1973}).

\bibitem{1992ApJ...397L..87M}
\bibinfo{author}{{Munari}, U.} \& \bibinfo{author}{{Renzini}, A.}
\newblock \bibinfo{title}{{Are symbiotic stars the precursors of type IA
  supernovae?}}
\newblock \emph{\bibinfo{journal}{\apjl}} \textbf{\bibinfo{volume}{397}},
  \bibinfo{pages}{L87--L90} (\bibinfo{year}{1992}).

\bibitem{1992A&A...262...97V}
\bibinfo{author}{{van den Heuvel}, E.~P.~J.}, \bibinfo{author}{{Bhattacharya},
  D.}, \bibinfo{author}{{Nomoto}, K.} \& \bibinfo{author}{{Rappaport}, S.~A.}
\newblock \bibinfo{title}{{Accreting white dwarf models for CAL 83, CAL 87 and
  other ultrasoft X-ray sources in the LMC}}.
\newblock \emph{\bibinfo{journal}{\aap}} \textbf{\bibinfo{volume}{262}},
  \bibinfo{pages}{97--105} (\bibinfo{year}{1992}).

\bibitem{2010A&A...523A...3L}
\bibinfo{author}{{Liu}, W.-M.}, \bibinfo{author}{{Chen}, W.-C.},
  \bibinfo{author}{{Wang}, B.} \& \bibinfo{author}{{Han}, Z.~W.}
\newblock \bibinfo{title}{{Helium-star evolutionary channel to
  super-Chandrasekhar mass type Ia supernovae}}.
\newblock \emph{\bibinfo{journal}{\aap}} \textbf{\bibinfo{volume}{523}},
  \bibinfo{pages}{A3} (\bibinfo{year}{2010}).

\bibitem{wlb+06}
\bibinfo{author}{Wizinowich, P.~L.} \emph{et~al.}
\newblock \bibinfo{title}{{The W. M. Keck Observatory Laser Guide Star Adaptive
  Optics System: Overview}}.
\newblock \emph{\bibinfo{journal}{\pasp}} \textbf{\bibinfo{volume}{118}},
  \bibinfo{pages}{297} (\bibinfo{year}{2006}).

\bibitem{2001ApJ...558..323H}
\bibinfo{author}{{Hachisu}, I.} \& \bibinfo{author}{{Kato}, M.}
\newblock \bibinfo{title}{{Recurrent Novae as a Progenitor System of Type Ia
  Supernovae. I. RS Ophiuchi Subclass: Systems with a Red Giant Companion}}.
\newblock \emph{\bibinfo{journal}{\apj}} \textbf{\bibinfo{volume}{558}},
  \bibinfo{pages}{323--350} (\bibinfo{year}{2001}).

\bibitem{2008ApJ...684.1366K}
\bibinfo{author}{{Kato}, M.}, \bibinfo{author}{{Hachisu}, I.},
  \bibinfo{author}{{Kiyota}, S.} \& \bibinfo{author}{{Saio}, H.}
\newblock \bibinfo{title}{{Helium Nova on a Very Massive White Dwarf: A Revised
  Light-Curve Model of V445 Puppis (2000)}}.
\newblock \emph{\bibinfo{journal}{\apj}} \textbf{\bibinfo{volume}{684}},
  \bibinfo{pages}{1366--1373} (\bibinfo{year}{2008}).

\bibitem{2001MNRAS.327.1323T}
\bibinfo{author}{{Thoroughgood}, T.~D.}, \bibinfo{author}{{Dhillon}, V.~S.},
  \bibinfo{author}{{Littlefair}, S.~P.}, \bibinfo{author}{{Marsh}, T.~R.} \&
  \bibinfo{author}{{Smith}, D.~A.}
\newblock \bibinfo{title}{{The mass of the white dwarf in the recurrent nova U
  Scorpii}}.
\newblock \emph{\bibinfo{journal}{\mnras}} \textbf{\bibinfo{volume}{327}},
  \bibinfo{pages}{1323--1333} (\bibinfo{year}{2001}).
\newblock \eprint{arXiv:astro-ph/0107477}.

\bibitem{2010ApJS..187..275S}
\bibinfo{author}{{Schaefer}, B.~E.}
\newblock \bibinfo{title}{{Comprehensive Photometric Histories of All Known
  Galactic Recurrent Novae}}.
\newblock \emph{\bibinfo{journal}{\apjs}} \textbf{\bibinfo{volume}{187}},
  \bibinfo{pages}{275--373} (\bibinfo{year}{2010}).

\bibitem{g00}
\bibinfo{author}{{Greiner}, J.}
\newblock \bibinfo{title}{{Catalog of supersoft X-ray sources}}.
\newblock \emph{\bibinfo{journal}{New Ast.}} \textbf{\bibinfo{volume}{5}},
  \bibinfo{pages}{137--141} (\bibinfo{year}{2000}).
\newblock \eprint{arXiv:astro-ph/0005238}.

\bibitem{d10a}
\bibinfo{author}{{Di Stefano}, R.}
\newblock \bibinfo{title}{{The Progenitors of Type Ia Supernovae. I. Are they
  Supersoft Sources?}}
\newblock \emph{\bibinfo{journal}{\apj}} \textbf{\bibinfo{volume}{712}},
  \bibinfo{pages}{728--733} (\bibinfo{year}{2010}).
\newblock \eprint{0912.0757}.

\bibitem{d10b}
\bibinfo{author}{{Di Stefano}, R.}
\newblock \bibinfo{title}{{The Progenitors of Type Ia Supernovae. II. Are they
  Double-degenerate Binaries? The Symbiotic Channel}}.
\newblock \emph{\bibinfo{journal}{\apj}} \textbf{\bibinfo{volume}{719}},
  \bibinfo{pages}{474--482} (\bibinfo{year}{2010}).
\newblock \eprint{1004.1193}.

\bibitem{ypr07}
\bibinfo{author}{{Yoon}, S.-C.}, \bibinfo{author}{{Podsiadlowski}, P.} \&
  \bibinfo{author}{{Rosswog}, S.}
\newblock \bibinfo{title}{{Remnant evolution after a carbon-oxygen white dwarf
  merger}}.
\newblock \emph{\bibinfo{journal}{\mnras}} \textbf{\bibinfo{volume}{380}},
  \bibinfo{pages}{933--948} (\bibinfo{year}{2007}).
\newblock \eprint{0704.0297}.

\bibitem{2011arXiv1108.4036S}
\bibinfo{author}{{Shen}, K.~J.}, \bibinfo{author}{{Bildsten}, L.},
  \bibinfo{author}{{Kasen}, D.} \& \bibinfo{author}{{Quataert}, E.}
\newblock \bibinfo{title}{{The Long-Term Evolution of Double White Dwarf
  Mergers}}.
\newblock \emph{\bibinfo{journal}{ArXiv e-prints}}  (\bibinfo{year}{2011}).
\newblock \eprint{1108.4036}.

\bibitem{1999ApJ...519..314H}
\bibinfo{author}{{Hachisu}, I.}, \bibinfo{author}{{Kato}, M.},
  \bibinfo{author}{{Nomoto}, K.} \& \bibinfo{author}{{Umeda}, H.}
\newblock \bibinfo{title}{{A New Evolutionary Path to Type IA Supernovae: A
  Helium-rich Supersoft X-Ray Source Channel}}.
\newblock \emph{\bibinfo{journal}{\apj}} \textbf{\bibinfo{volume}{519}},
  \bibinfo{pages}{314--323} (\bibinfo{year}{1999}).
\newblock \eprint{arXiv:astro-ph/9902303}.

\bibitem{2010AJ....140.1158T}
\bibinfo{author}{{Torres}, G.}
\newblock \bibinfo{title}{{On the Use of Empirical Bolometric Corrections for
  Stars}}.
\newblock \emph{\bibinfo{journal}{\aj}} \textbf{\bibinfo{volume}{140}},
  \bibinfo{pages}{1158--1162} (\bibinfo{year}{2010}).

\end{thebibliography}
\end{document}